\documentclass[twocolumn,epjc3]{svjour3}
\smartqed  
\RequirePackage{graphicx}
\usepackage{color}

\begin{document}

\title{ Horizon structure of rotating Einstein-Born-Infeld black holes and shadow}
\author{
Farruh Atamurotov\thanksref{1,a,b} \and
Sushant~G.~Ghosh\thanksref{2,c,d} \and
Bobomurat Ahmedov\thanksref{3,a,e} \and}
\institute{Institute of Nuclear Physics, Ulughbek, Tashkent 100214, Uzbekistan\label{a}\and Inha University in Tashkent, Tashkent 100170, Uzbekistan\label{b}\and
Centre for Theoretical Physics, Jamia Millia
Islamia, New Delhi 110025, India\label{c}\and
 Astrophysics and Cosmology
Research Unit, School of Mathematical Sciences, University of
Kwa-Zulu-Natal, Private Bag 54001, Durban 4000, South Africa\label{d}
\and Ulugh Beg Astronomical Institute, Astronomicheskaya 33, Tashkent 100052, Uzbekistan\label{e}}
\thankstext{1}{\emph{e-mail:} farruh@astrin.uz, fatamurotov@gmail.com}
\thankstext{2}{\emph{e-mail:} sghosh2@jmi.ac.in,sgghosh@gmail.com}
\thankstext{3}{\emph{e-mail:} ahmedov@astrin.uz}

\maketitle
\begin{abstract}
We investigate the horizon structure  of the rotating
Einstein-Born-Infeld solution which goes over to the Einstein-Maxwell's
Kerr-Newman solution as the Born-Infeld parameter goes to infinity
($\beta \rightarrow \infty$). We find that for a given $\beta$,
mass $M$ and charge $Q$, there exist critical spinning parameter $a_{E}$ and $r_{H}^{E}$, which
corresponds to an  extremal Einstein-Born-Infeld black hole with
degenerate horizons, and $a_{E}$ decreases and $r_{H}^{E}$ increases
with increase in the Born-Infeld parameter $\beta$. While $a<a_{E}$ describe a  non-extremal
Einstein-Born-Infeld black hole with outer and inner horizons. Similarly, the effect of $\beta$ on infinite redshift surface and in turn on ergoregion is also included. It is
well known that a black hole can cast a shadow as an optical
appearance due to its strong gravitational field. We also
investigate the shadow cast by the non-rotating ($a=0$)
Einstein-Born-Infeld black hole and  demonstrate that the null
geodesic equations can be integrated that allows us to investigate
the shadow cast by a black hole which is found to be a dark zone
covered by a circle.  Interestingly, the shadow of the Einstein-Born-Infeld black
hole is slightly smaller than for the Reissner-Nordstrom black hole. F
urther, the shadow is concentric circles whose radius decreases with increase in
value of parameter $\beta$.
\end{abstract}

\section{Introduction}
In Maxwell's electromagnetic field theory, the field of a point-like charge is singular at
the charge position and hence it has infinite self-energy. To
overcome this problem in classical electrodynamics, the non-linear electromagnetic field has been proposed by
Born and Infeld \cite{bi}, with main motivation, to resolve
self-energy problem by imposing a maximum strength of the
electromagnetic field.   In this theory  the electric field of a
point charge is regular at the origin and this non-linear theory for
the electromagnetic field was able to tone down the infinite self
energy of the point-like charged particle. Later, Hoffmann
\cite{Hoffmann} coupled has general relativity with Born-Infeld
electrodynamics to obtain a spherically symmetric solutions for the
gravitational field of an electrically charged object. Remarkably, after a long
discard, the Born-Infeld theory  made a come back to the stage in
the context of more modern developments, which is mainly due to the
interest in non-linear electrodynamics in the context of low energy
string theory, in which Born-Infeld type actions appeared
\cite{string}. Indeed,  the low energy effective action in an open
superstring in loop calculations lead to Born-Infeld type actions
\cite{string1}. These important features of the Born-Infeld theory,
together with its corrective properties concerning singularities,
further motivate to search for gravitational analogues of this
theory in the past \cite{past}, and also interesting measures have
been taken to get the spherically symmetric solutions \cite{bhbi}.
The thermodynamic properties and causal structure of the
Einstein-Born-Infeld black holes drastically differ from that of
the classical Reissner-Nordstrom black holes. Indeed,  it turns out
that the Einstein-Born-Infeld black hole singularity is weaker than
that of Reissner-Nordstrom black hole. Further properties of these
black holes, including motion of the test particles has been also
addressed \cite{lmc}. It is worthwhile to mention that Kerr
\cite{kerr} and Kerr-Newman metrics \cite{nja} are undoubtedly the
most significant exact solutions in the general relativity, which
represent rotating black hole that can arise as  the final fate of
gravitational collapse. The generalization of the spherically
symmetric Einstein-Born-Infeld black hole in the rotating case,
Kerr-Newman like solution, was studied by Lombardo \cite{CiriloLombardo:2004qw}. In
particular, it is demonstrated \cite{bi} that the rotating
Einstein-Born-Infeld solutions can be derived starting from the
corresponding exact spherically symmetric solutions \cite{Hoffmann}
by a complex coordinate transformation previously developed by
Newman and Janis \cite{nja}. The rotating Einstein-Born-Infeld black
hole  metrics are axisymmetric, asymptotically flat and depend on
the mass, charge and spin of the black hole as well as on a
Born-Infeld parameter ($\beta$) that measure potential deviations
from the Kerr metric Kerr-Newman metrics. The rotating
Einstein-Born-Infeld  metric includes the Kerr-Newman metric as the
special case if this deviation parameter diverges ($\beta
\rightarrow \infty$) as well as the Kerr metric when this parameter
vanishes ($\beta =0$).  In this paper, we carry out detailed
analysis of the horizon structure of rotating Einstein-Born-Infeld
black hole and explicitly manifest the impact that the parameter
$\beta$ makes. Recently, horizon structure has been studied
for various space-time geometris, see, e.g.,~\cite{Shoom}.  
We also investigate the apparent shape of
non-rotating Einstein-Born-Infeld black hole to visualize the shape
of the shadow and compare the results with the images for the
corresponding Reissner-Nordstrom black hole.  In spite of the fact
that's a black hole is invisible, its shadow can be observed if it is
in front of a bright background~\cite{Lu} as the result of the
gravitational lensing effect, see, e.g.,~\cite{vih1,schee1}. The
photons that cross the event horizon, due to strong gravity, are
removed from the observable universe which lead  to a shadow
(silhouette) imprinted by a black hole on the bright emission that
exists in its vicinity. So far the shadow of the compact
gravitational objects in the different cases have been extensively
studied, see, e.g.,~\cite{Fara}. Furthermore new general formalism
to describe the shadow of black hole as an arbitrary polar curve
expressed in terms of a Legendre expansion is  developed in the recent paper ~\cite{Rezz}.
The organization of the paper is as follows: In the Section II, we study the structure and location of an event horizon and the infinite redshift surface of the rotating Einstein-Born-Infeld black holes.
We have also discussed the particle motion around the rotating
Born-Infeld black hole in the Section II that helped us to investigate  the shadow of the black hole in the Section III.  In Section IV emission energy is analysed of rotating Einstein-Born-Infeld black holes.
Finally, in Section V, we conclude by summarizing the main results.
We use units which fix the speed of light and the gravitational constant via $G = c = 1$, and use the metric signature
($-,\;+,\;+,\;+$).
\section{Rotating Einstein-Born-Infeld black hole}
The action for the gravitational field coupled to a  nonlinear
Born-Infeld electrodynamics (or a Einstein-Born-Infeld action) in $
(3+1) $ dimensions reads \cite{bi,Hoffmann}
\begin{equation}\label{action}
S= \int d^{4}x\sqrt{-g}\left[ \frac{R}{16\pi
G}+\mathcal{L}(\mathcal{F})\right]\ ,
\end{equation}
where $R$ is scalar curvature, $g \equiv \det|g_{\mu \nu}|$ and $
\mathcal{L}(\mathcal{F}) $ is given by
\begin{equation}
\mathcal{L}(\mathcal{F})=\frac{\beta^{2}}{4\pi G}\left(
1-\sqrt{1+\frac{2\mathcal{F}}{\beta^{2}}}\right),
\end{equation}
with $ \mathcal{F}=\frac{1}{4} F_{\mu\nu}F^{\mu\nu}$, $F_{\mu\nu}$
denotes  the electromagnetic field tensor. Here  $ \beta^2 $ is the
Born-Infeld parameter being equal to the maximum value of electromagnetic field intensity
and has a dimension of $[length]^{-2}$.  Eq.~(\ref{action}) leads to
the Einstein field equations
\begin{equation}\label{efe}
R_{\mu \nu}-\frac{1}{2}g_{\mu \nu}R= \kappa T_{\mu \nu}\ ,
\end{equation}
and electromagnetic field equation
\begin{equation}\label{emf}
\nabla_{\mu}(F^{\mu \nu} \mathcal{L,_F})=0\ .
\end{equation}
The energy momentum tensor is
\begin{equation}\label{emt}
T_{\mu \nu}=\mathcal{L}g_{\mu \nu}-F_{\mu \sigma}F_{\nu}^{\sigma}\ ,
\end{equation}
where $\mathcal{L,_F}$ denotes partial derivative of $\mathcal{L}$
with respect to $F$.
\begin{figure*}
    \begin{tabular}{|c|c|}
    \hline
        \includegraphics[scale=0.8]{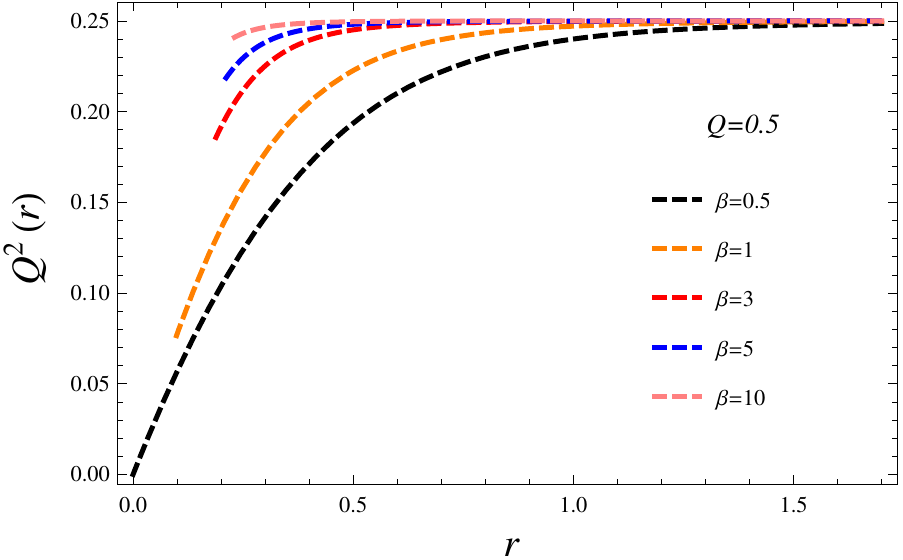}&
        \includegraphics[scale=0.8]{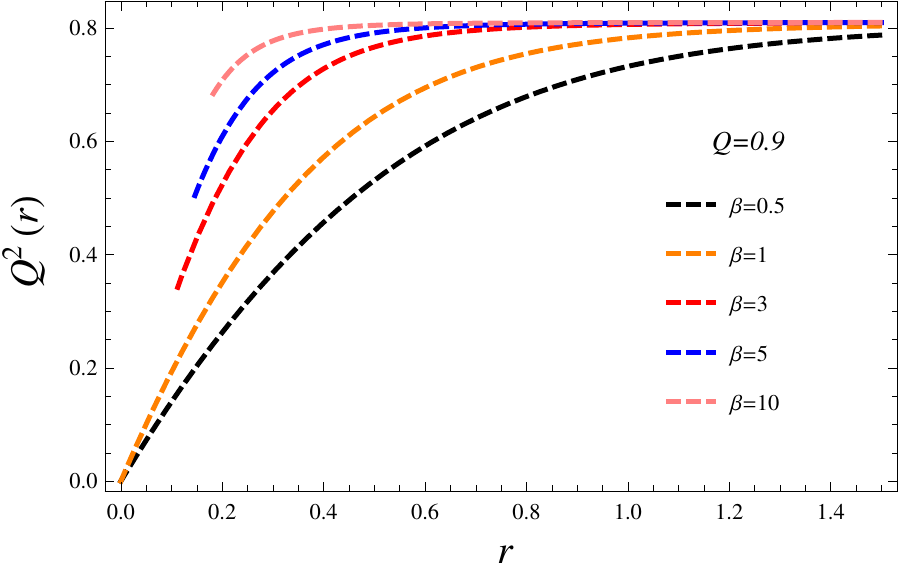} \\
         \hline
    \end{tabular}
    \caption{Plots showing the dependence of the  square of the electric charge $Q^2(r)$ from the radial coordinate $r$ for
        different values of Born-Infeld parameter $\beta$. Left panel is for $Q=0.5$ and right panel is for $Q=0.9$ in asymptotics.}\label{q}
\end{figure*}
\begin{figure*}
    \begin{tabular}{|c|c|}
    \hline
        \includegraphics[scale=0.8]{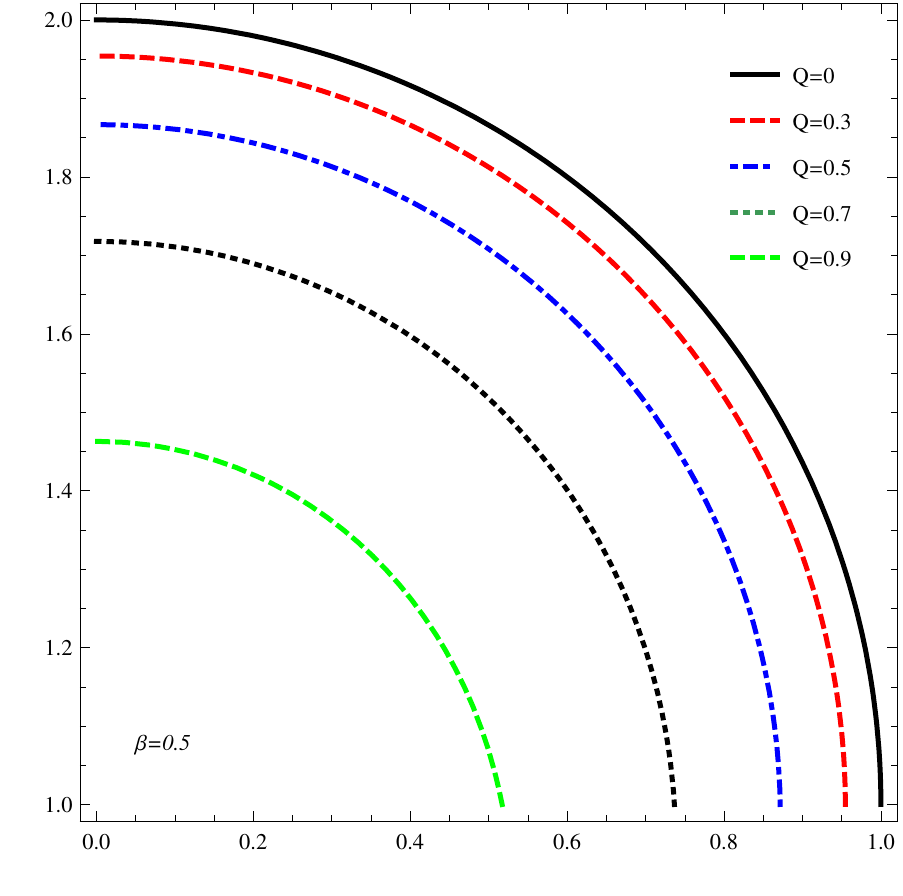}&
        \includegraphics[scale=0.8]{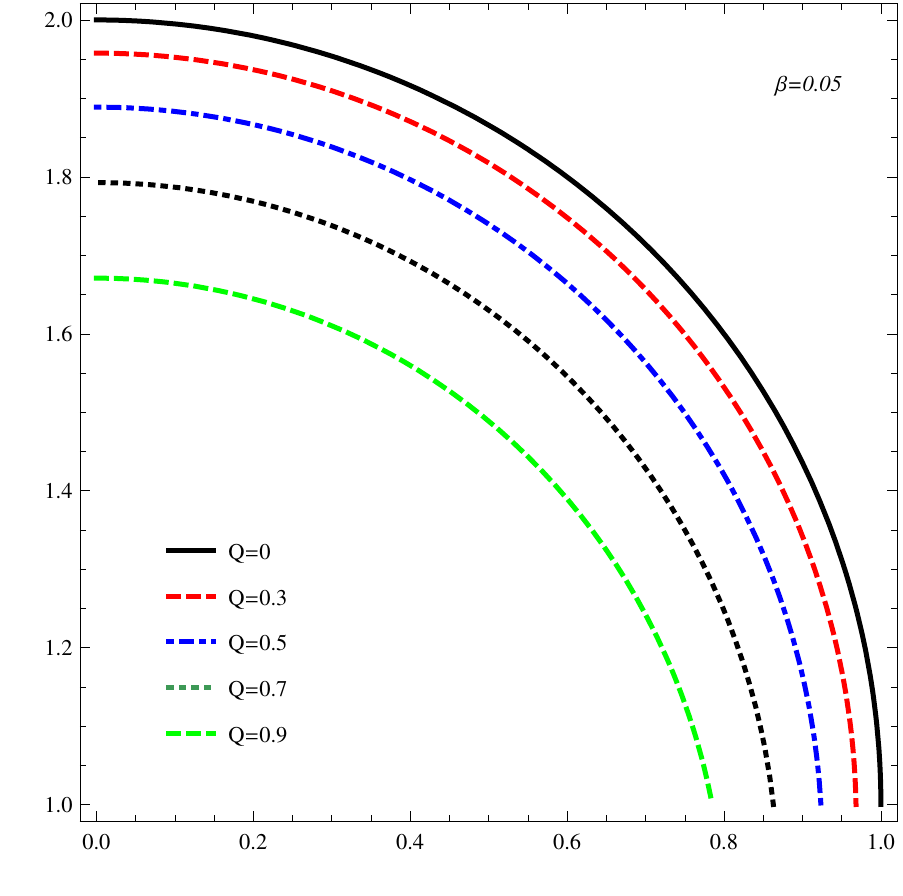} \\
         \hline
    \end{tabular}
    \caption{ The rotation parameter $a$ dependence of the radial coordinate $r$ for the different value of electric charge $Q$ and Born-Infeld parameter $\beta$. The lines separate the region of black holes with naked singularity ones. The left panel is for Born-Infeld parameter $\beta=0.5$ and the right panel is for Born-Infeld parameter $\beta=0.05$.
    }\label{hq}
\end{figure*}
\begin{figure*}
    \begin{tabular}{|c|c|}
    \hline
        \includegraphics[scale=0.8]{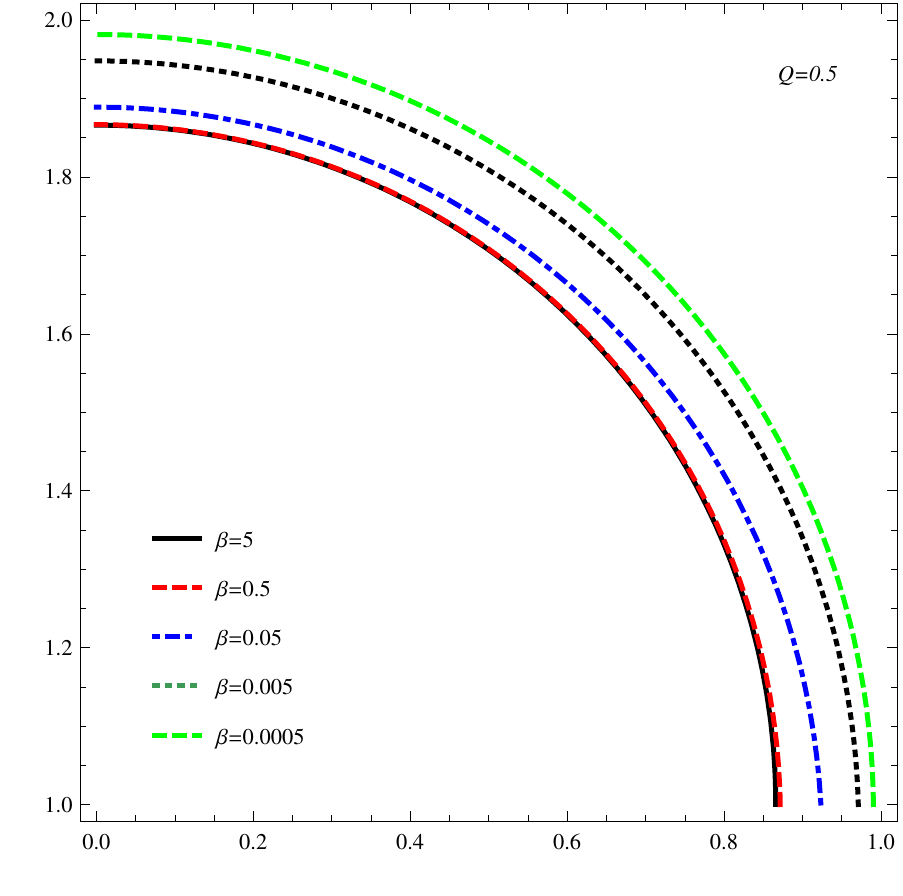}&
        \includegraphics[scale=0.8]{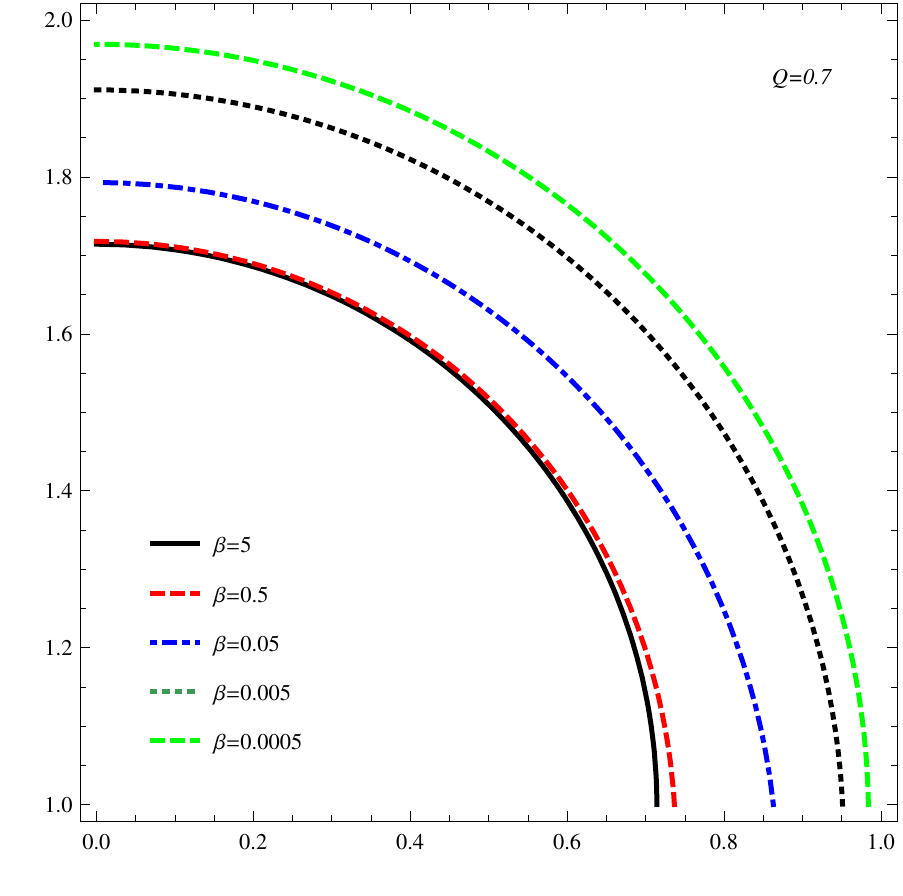} \\
         \hline
    \end{tabular}
    \caption{ The rotation parameter $a$ dependence of the radial coordinate $r$ for the different value of electric charge $Q$ and Born-Infeld parameter $\beta$. The lines separate the region of black holes with naked singularity ones. The left panel is for electric charge $Q=0.5$ and the right panel is for electric charge $Q=0.7$. }\label{hb}
\end{figure*}
\begin{figure*}
    \begin{tabular}{|c|c|}
    \hline
        \includegraphics[scale=0.8]{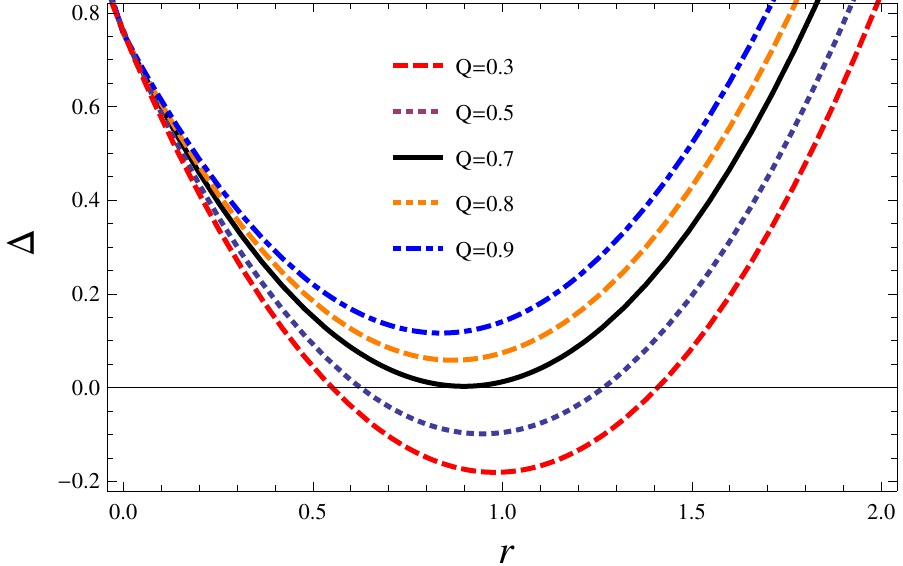}&
        \includegraphics[scale=0.8]{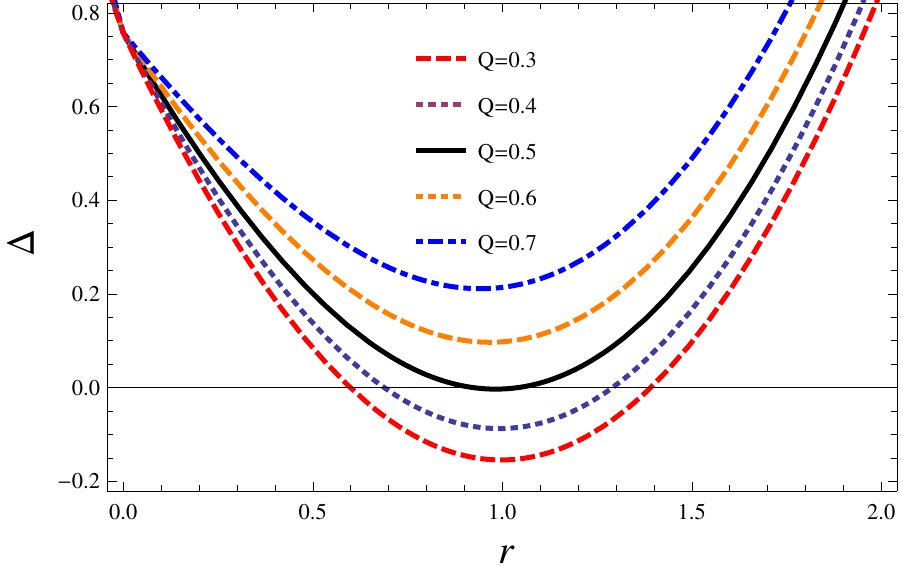} \\
         \hline \\
        \includegraphics[scale=0.8]{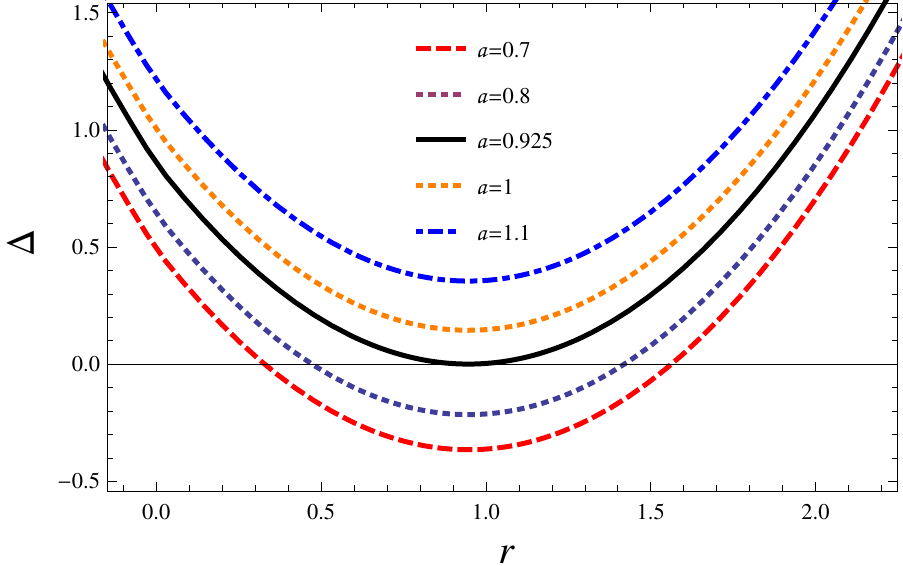} &
        \includegraphics[scale=0.8]{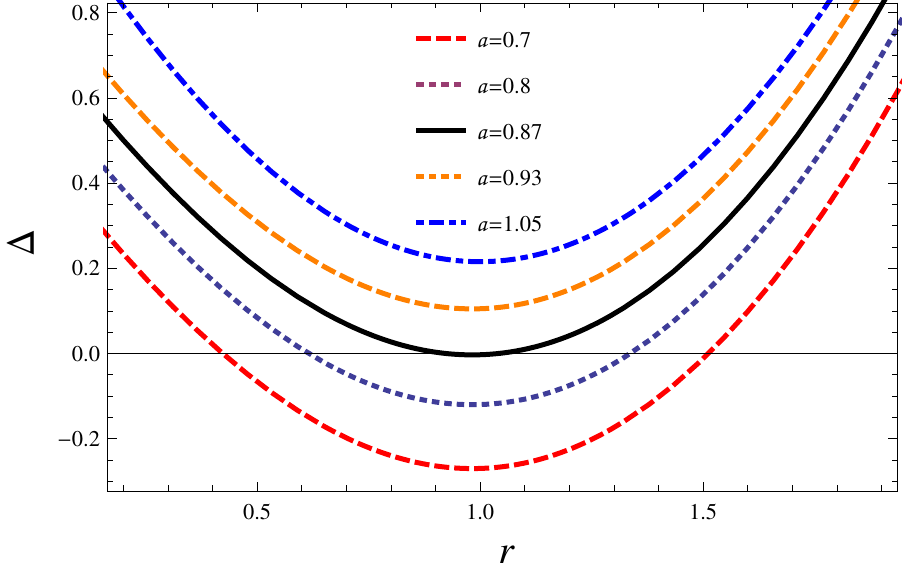}   \\
         \hline
    \end{tabular}
    \caption{Plots showing the radial dependence of $\Delta$ for the
        different values of Born-Infeld parameter $\beta$, electric charge $Q$ and rotation parameter $a$ (with $M=1$). Top, left panel is for $a=0.87$ and $\beta=0.05$. Top, right panel is for $a=0.87$ and $\beta=0.5$. Bottom, left panel is for $Q=0.5$ and $\beta=0.05$. Bottom, right panel is for $Q=0.5$ and $\beta=0.5$. }\label{ehf}
\end{figure*}

The gravitational  field of a static and spherically symmetric
compact object with mass $M$ and a non-linear electromagnetic source
in the Einstein-Born-Infeld theory has first been investigated by
Hoffmann \cite{Hoffmann} and the space-time metric is
\cite{Hoffmann,Gibbons}
\begin{eqnarray}\label{hg}
& & ds^{2}=-\left[ 1-\frac{2GM}{r}+\frac{Q^{2}\left( r\right) }{r^{2}}\right]
dt^{2} \nonumber \\ & &  +\left[ 1-\frac{2GM}{r}+\frac{Q^{2}\left( r\right) }{r^{2}}\right]
^{-1}dr^{2} +r^{2}\left( d\theta ^{2}+\sin ^{2}\theta \ d\varphi ^{2}\right)\ ,\nonumber\\
\end{eqnarray}
with the square of electric charge
\begin{eqnarray}
Q^2(r) & & = \frac{2\beta^{2}r^{4}}{3}\left( 1-\sqrt{1+\zeta^2(r)}\right)
\nonumber \\ & & +\frac{4Q^{2}}{3} F\left( \frac{1}{4},\frac{1}{2},\frac{5}{4},-\zeta^2(r)\right)\ ,
\end{eqnarray}
where $ F $ is a Gauss hypergeometric function \cite{hyper} and new notation $\zeta^2(r)=Q^2/(\beta^2 r^4)$ is introduced.
From the radial dependence of $Q^2(r)$ plotted in Fig.\ref{q} one can see
its strong dependence from  $\beta$ parameter near to the center of the black hole.
The rotating counterpart  of the Einstein-Born-Infeld black hole has been
obtained in \cite{CiriloLombardo:2004qw}. The gravitational field of
rotating Einstein-Born-Infeld black hole spacetime
 is described by the metric which in the
Boyer-Lindquist coordinates is given by ~\cite{CiriloLombardo:2004qw}
\begin{eqnarray}\label{bi}
ds^2 && = \frac{\Delta - a^2 \sin^2 \theta}{\rho^2} dt^2 - \frac{\rho^2}{\Delta }  \, dr^2    \nonumber \\
&&
 +  2 a \sin^2 \theta \left(1 - \frac{\Delta - a^2 \sin^2 \theta}{\rho^2} \right) dt \, d \phi
 - \rho^2 \, d \theta^2
 \nonumber \\
&&
 -  \, \sin ^2 \theta  \left[ \rho^2 + a^2 \sin^2 \theta \left(2 - \frac{\Delta - a^2 \sin^2\theta}{\rho^2}\right)   \right]    d \phi^2\ ,\nonumber\\
  \label{SPK2}
\end{eqnarray}
with \begin{equation} \Delta = r^{2}-2GMr+Q^{2}\left( r\right)
+a^{2},\mbox{and}\; \rho^2=r^2+a^2\cos^2\theta.
\end{equation}
The parameters $a$, $M$, $Q$ and $\beta$ are, respectively correspond to
rotation,  mass, the electric charge and the Born-Infeld parameter. We let the parameters $Q$
and $\beta$ to be positive.   In the limit $ \beta \rightarrow \infty
$ (or $Q(r)=Q$) and $ Q\neq0 $, one obtains the corresponding
solution for Kerr-Newman black hole, while one has Kerr black hole
\cite{kerr} when $\beta \rightarrow 0$. The metric (\ref{bi}) is a
rotating charged black hole which generalizes the standard
Kerr-Newman black hole and we call it as the rotating
Einstein-Born-Infeld black hole. The non-rotating case, $a=0$,
corresponds to the metric of the static Einstein-Born-Infeld black
hole obtained by Hoffmann in \cite{Hoffmann}. The metric (\ref{bi})
has curvature singularity at  the set of  points, where $\rho=0$ and
$M=Q\neq 0$. For $a\neq0$, it corresponds to a ring with radius $a$,
in the equatorial plane $\theta=\pi/2$ and hence termed as a ring
singularity.
The properties of the rotating Einstein-Born-Infeld metric
(\ref{bi}) are  similar to that of the general relativity counterpart
Kerr-Newman black hole. We first show that, it is possible to get
certain range of values of $a$, $M$ and $Q$, the metric (\ref{bi})
is a black hole.  The metric (\ref{bi}), like the Kerr-Newman one, is
singular at $\Delta=0$ and it admits two horizons like surfaces, viz.,
thee static limit surface and the event horizon. Here, we shall look for
these two surfaces for the rotating Einstein-Born-Infeld metric
(\ref{bi}) and discuss effect of the nonlinear parameter $\beta$.
The horizons of the Einstein-Born-Infeld black hole (\ref{bi}) are
dependent on parameters $M,a,Q$ and $\beta$, and  are calculated by
equating the $g^{rr}$ component of the metric (\ref{bi}) to zero,
i.e.,
\begin{equation}\label{eh}
\Delta = r^{2}-2GMr+Q^{2}\left( r\right) +a^{2}=0,
\end{equation}
which depends on $Q(r)$ a function of $r$, and is different from the
Kerr-Newman black hole, where $Q$ is just a constant. The solution Eq.~(\ref{eh}) can have either
no roots (naked singularity), two roots (horizons) depending
on the values of these parameters. It is difficult to solve the
Eq.~(\ref{eh}) analytically and hence we approach for numerical solutions.
It is seen that Eq.~(\ref{eh}) admits two horizons $r_{EH}^{-}$ and
$r_{EH}^{+}$ for suitable choice of parameters, which corresponds to
two positive roots of Eq.~(\ref{eh}), with $r_{EH}^{+}$ determines
the event horizons and $r_{EH}^{-}$ the Cauchy horizon. Further, it
is worthwhile to mention that one can set parameters when
$r_{EH}^{-}$ and $r_{EH}^{+}$  are equal and we have an extremal
black hole. We have plotted the event horizons in
Fig.~\ref{hq}-\ref{hb}  for different values of mass $M$, charge
$Q$, parameter $\beta$ and spinning parameter $a$. Like, the Kerr-Newman
black hole, the rotating spacetime  (\ref{bi}) has two horizons,
viz., the Cauchy horizon and the event horizon. The  figures reveals
that there exists set of values of parameters for which we have two
horizons, i.e., a  black hole with both inner and outer horizons.
One can also find values of parameters for which one get an extremal
black hole where the two horizons coincide. The region between
the static limit surface and the event horizon is termed as quantum
ergosphere, where  it is possible to enter and leave again, and the
object moves in the direction of the spin of black hole.
\begin{figure*}
    \begin{tabular}{|c|c|}
    \hline
        \includegraphics[scale=0.8]{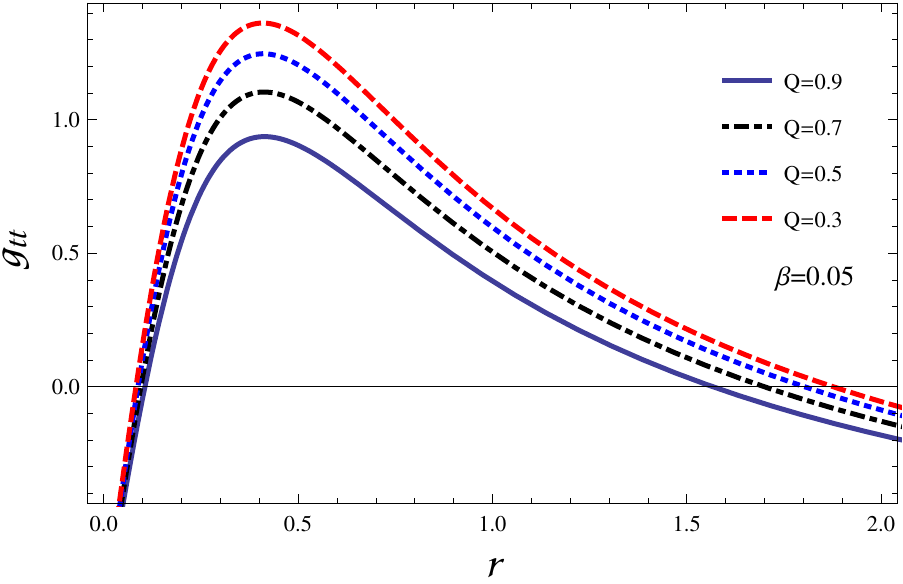}&
        \includegraphics[scale=0.8]{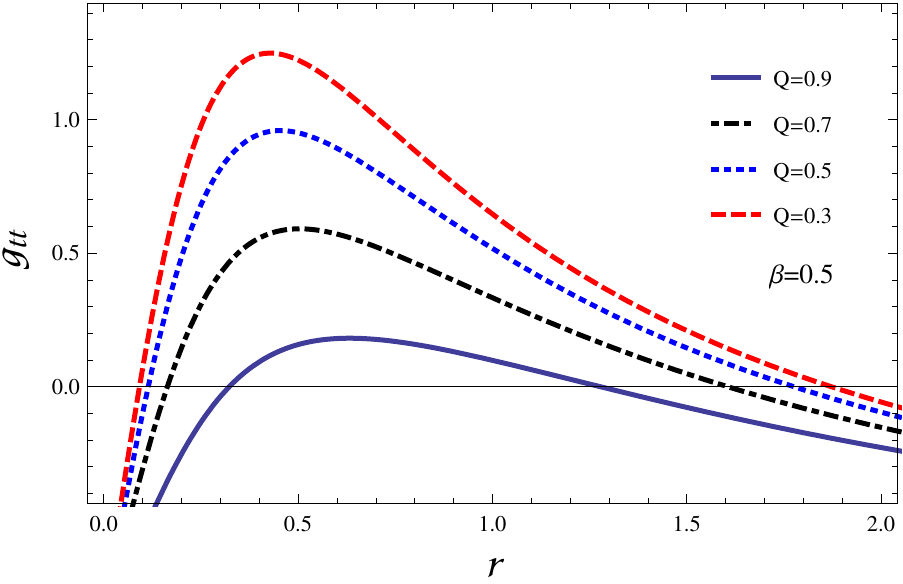} \\
         \hline \\
        \includegraphics[scale=0.8]{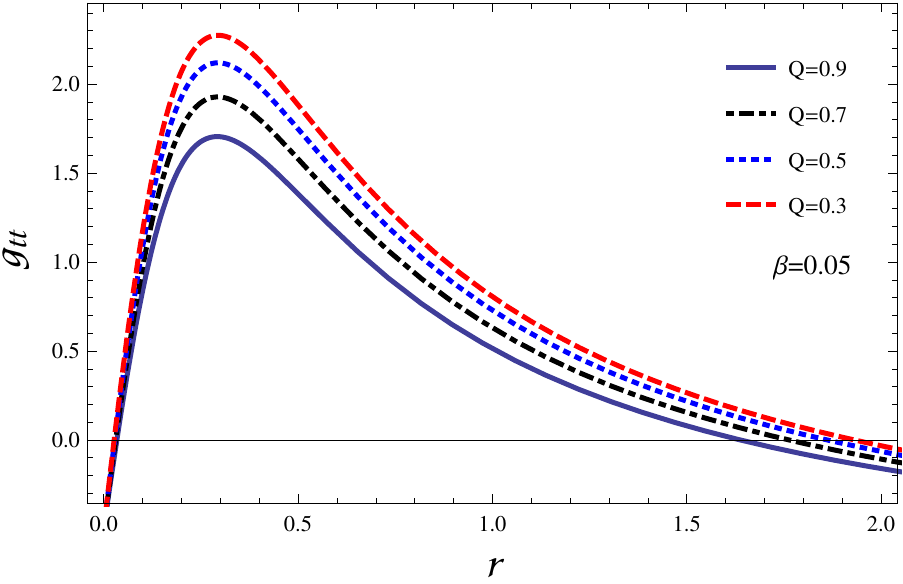} &
        \includegraphics[scale=0.8]{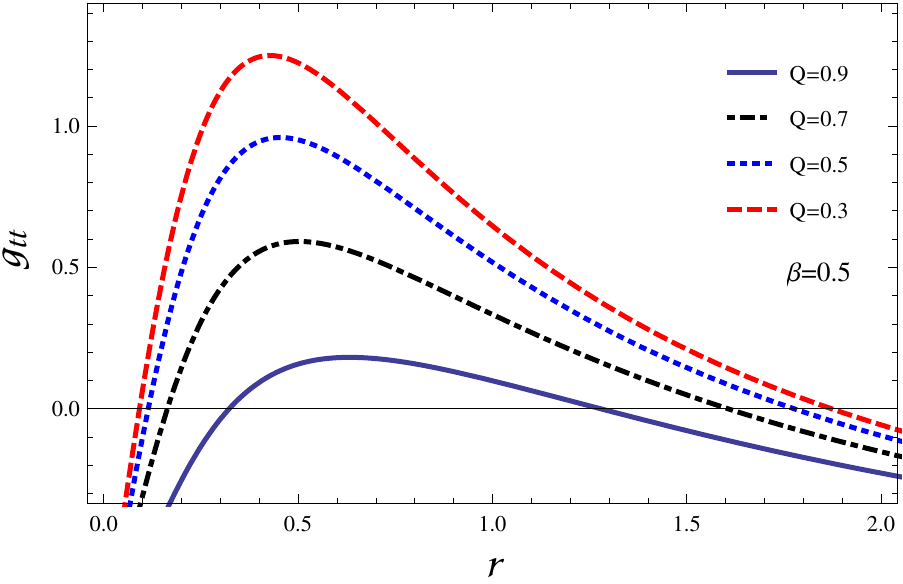}   \\
         \hline
    \end{tabular}
    \caption{Plots showing the radial dependence of $g_{tt}$ component of metric tensor for
        the different values of Born-Infeld parameter $\beta$ and electric charge $Q$ (with $M=1$). Top, left panel is for $\beta=0.05$, $a=0.45$ and $\alpha=\pi/6$. Top, right panel is for $\beta=0.5$, $a=0.45$ and $\alpha=\pi/6$. Bottom, left panel is for $\beta=0.05$, $a=0.45$ and $\alpha=\pi/3$. Bottom, right panel is for $\beta=0.5$, $a=0.45$ and $\alpha=\pi/3$.}\label{gtt1}
\end{figure*}
\begin{figure*}
   \begin{tabular}{|c|c|}
    \hline
       \includegraphics[scale=0.8]{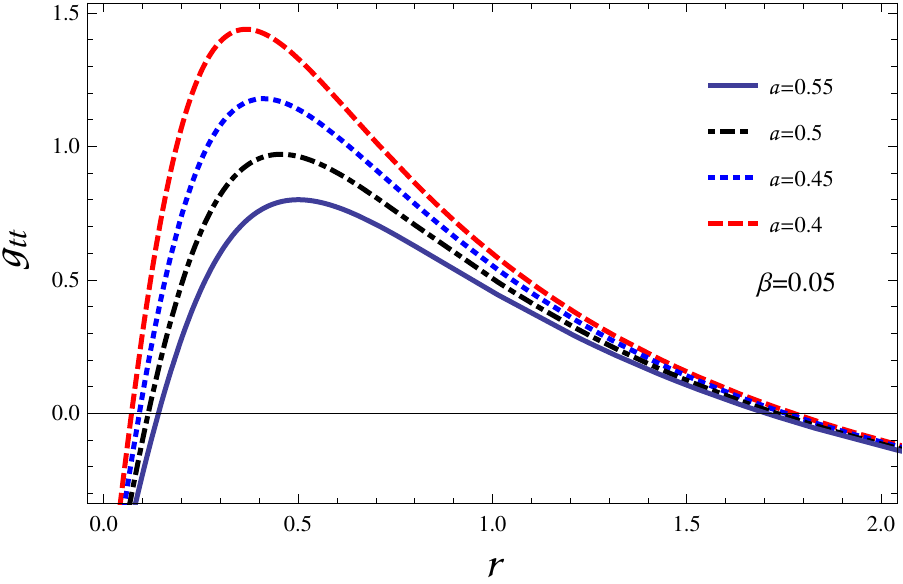}&
        \includegraphics[scale=0.8]{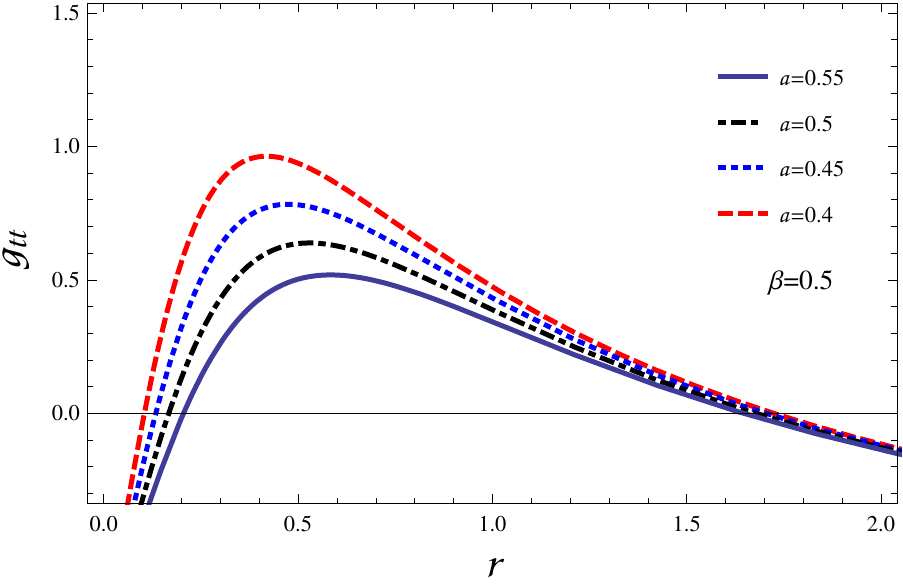} \\
       \hline \\
        \includegraphics[scale=0.8]{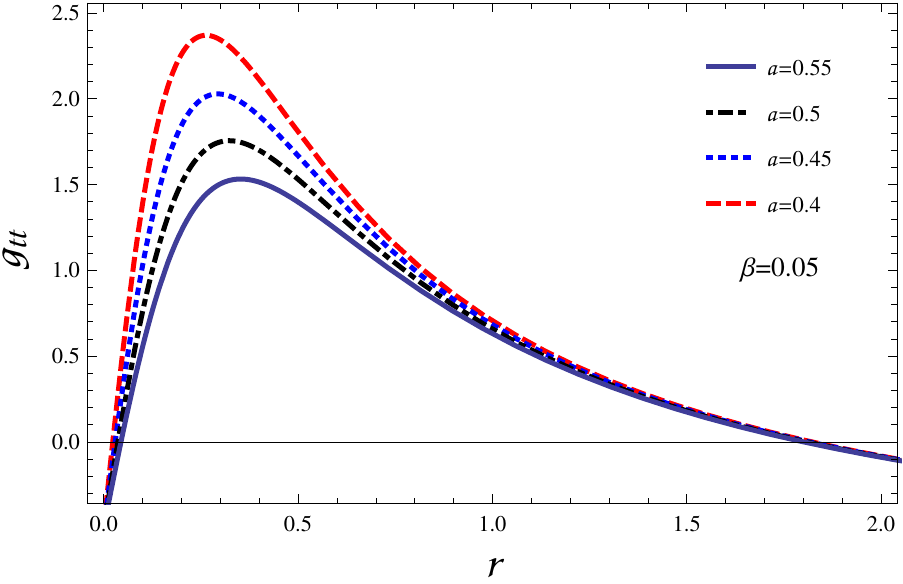} &
        \includegraphics[scale=0.8]{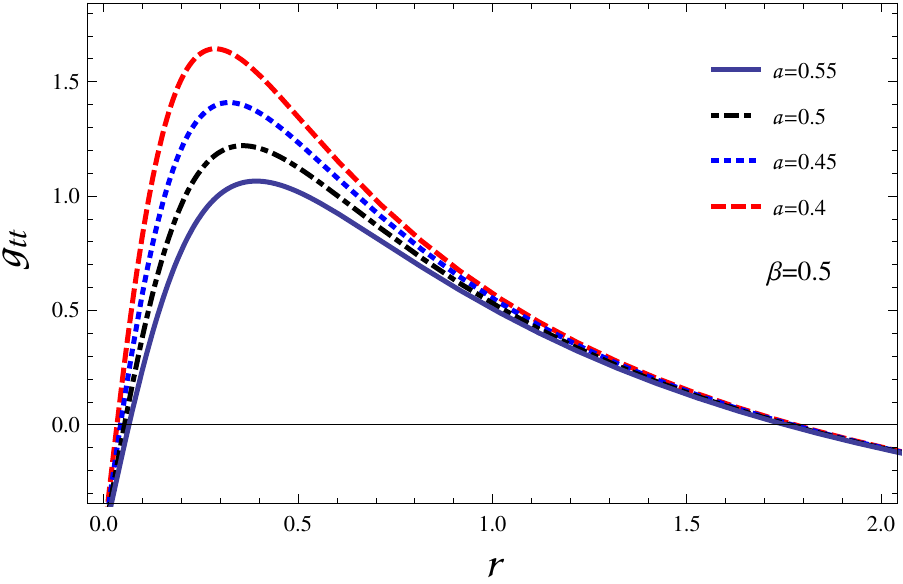}   \\
         \hline
    \end{tabular}
    \caption{Plots showing the radial dependence of $g_{tt}$ component of metric tensor for
        the different values of parameter $\beta$ and rotation parameter $a$ (with $M=1$).  Top, left panel is for $\beta=0.05$, $Q=0.6$ and $\alpha=\pi/6$. Top, right panel is for $\beta=0.5$, $Q=0.6$ and $\alpha=\pi/6$. Bottom, left panel is for $\beta=0.05$, $Q=0.6$ and $\alpha=\pi/3$. Bottom, right panel is for $\beta=0.5$, $Q=0.6$ and $\alpha=\pi/3$.}\label{gtt2}
\end{figure*}
We have numerically studied the horizon properties for nonzero
values of $a,\;\beta$ and $Q$ (cf. Fig.~\ref{ehf}) by  solving
Eq.~(\ref{eh}). It turns out that the Born-Infeld parameter $\beta$
makes a profound influence on the horizon structure when compared with the Kerr black hole.  We find that
for a given values of parameters $\beta,Q$, there exist extremal
value of $a=a_{E}$ and $r=r^{E}_{H}$ such that for $a<a_{E}$,
Eq.~(\ref{eh}) admits two positive roots , which corresponds to
respectively, a black hole has two horizons or  black hole with both
Cauchy and event horizons. We found no root at $a>a_{E}$ (naked
singularity) (see Fig.~\ref{ehf}), i.e., existence of a naked
singularity. Further, one can find values of parameters for which
these two horizons coincide and we get extremal black holes.
Similarly, we have shown that for given values of parameters
$a,\beta$, we get an extremal value of $Q=Q_E$, for which two
horizons coincide and we get extremal black holes as shown in
Fig.~\ref{ehf}. Interestingly, the value of $Q_E$ decreases with
increase in $\beta$.
\paragraph{Infinite red-shift surface or static limit surface.}
While in non-rotatig black hole, in general, the horizon is also the
surface where $g_{tt}$ changes sign, in rotating
Einstein-Born-Infeld, like Kerr-Newman, these surfaces do not
coincide. The location of infinite redshift surface or static limit
surface  requires the coefficient of $d{t}^2$ to vanish, i.e., it must satisfy
\begin{eqnarray}\label{sls}
 r^{2}-2GMr+Q^{2}\left(
r\right) +a^{2} \cos^2(\theta) = 0.
\end{eqnarray}
Eq.~(\ref{sls}) is solved numerically and the behavior of the static
limit  surface which is shown in Figs.~\ref{gtt1} and \ref{gtt2}. The
Einstein-Born-Infeld metric~(\ref{bi}) admits two static limit
surfaces $r_{SLS}^{-}$ and $r_{SLS}^{+}$ corresponding to two
positive roots of Eq.~(\ref{sls}) when the parameters $ M, Q, a,$
and $ \beta $ are chosen suitably (cf. Figs.~\ref{gtt1} and \ref{gtt2}).
Interestingly the radius of the  static limit surface decreases
with increase in the value of parameter $\beta$.  The static limit surface has
similar extremal behavior which is depicted in the Figs.~\ref{gtt1}
and \ref{gtt2}).   Like any other rotating black hole, there is a
region outside the outer horizon where $g_{tt}
> 0 $. The region, i.e. $r_{SLS}^{+} < r < r_{EH}^{+}$ is called ergoregion, and
its outer boundary $r = r_{SLS}^{+}$ is called the quantum
ergosphere.
\paragraph{Null geodesics in Einstein-Born-Infeld black hole space-time.}
 Next, we turn our attention to the study the geodesic of
a photon.
 We need to study the separability of the Hamilton-Jacobi equation
using  the approach due to Carter \cite{Carter68}. First, for generality we consider a
motion for a particle with mass $m_{0}$ falling in  the background
of a rotating Einstein-Born-Infeld black hole. The geodesic motion
for this black hole is determined by the following Hamilton-Jacobi
equations
\begin{equation}\label{hje}
\frac{\partial S}{\partial \tau} = -\frac{1}{2} g^{\mu\nu} \frac{\partial S}{\partial x^{\mu}} \frac{\partial S}{\partial x^{\nu}},
\end{equation}
where $\tau$ is an affine parameter along the geodesics, and $S$ is
the  Jacobi action. For this black hole background the Jacobi action
$ S $ can be separated as
\begin{equation}\label{hja}
S = \frac{1}{2} m_0^2 \tau -Et + L \phi + S_{r}(r) + S_{\theta}(\theta),
\end{equation}
where $ S_{r} $ and $ S_{\theta} $ are respectively functions of radial coordinate $ r
$ and angle $ \theta $. Like the Kerr space-time, rotating Born-Infeld
black hole also has two Killing vector fields due to the assumption
of stationarity and axisymmetry of the space-time, which in turn
guarantees the existence of two conserved quantities for a geodesic
motion, viz. the energy $ E $ and the axial component of the angular
momentum $ L $.  Thus, the constants $ m_0 $, $ E $, and $ L $
correspond to rest mass, conserved energy and rotation parameter
related through $m_{0}^2= -p_{\mu}p^{\mu}$, $E=-p_{t}$, and
$L=p_{\phi}$. Obviously for a photon null geodesic, we have
$m_{0}=0$, and from (\ref{hje}) we obtain the null geodesics in the
form of the first-order differential equations
\begin{eqnarray}
\rho^2\frac{dt}{d\tau}&=& \frac{r^2+a^2}{\Delta}\left[(r^2+a^2){\cal E} -a {\cal L}
\right]+\nonumber\\  \label{bibh1}
&&  a({\cal L} -a
{\cal E} \sin^2 \theta),
\\
\rho^2\frac{d\phi}{d\tau}&=&\frac{a}{\Delta }\left[(r^2+a^2){\cal E}
-a {\cal L} \right]+ \nonumber\\
&&(\frac{\cal
L}{\sin^2 \theta} -a {\cal E} ),  \label{bibh2}
\\
\rho^2\frac{dr}{d\tau}&=&\sqrt{\mathcal{R}},
 \label{bibh3}
\\
\rho^2\frac{d\theta}{d\tau}&=&\sqrt{\Theta}, \label{bibh4}
\end{eqnarray}
where the functions $\mathcal{R}(r)$ and $\Theta(\theta)$ are defined as
\begin{eqnarray}
\mathcal{R}&=&\left[(r^2+a^2){\cal E} -a {\cal L}
\right]^2-\Delta \left(\mathcal{K}+({\cal L} - a {\cal E})^2 \right), \
\\ \label{bibh5}
\Theta &=&\mathcal{K}+\cos^2\theta \left(a^2 {\cal E}^2
-\frac{{\cal L} ^{2}}{\sin^2\theta}\right). \nonumber \label{bibh6}
\end{eqnarray}
Thus, we find that Hamilton-Jacobi Eq.~(\ref{hje}), using
(\ref{hja}), is separable due existence of $ \mathcal{K} $ namely
Carter constant of separation.   The above equations govern the
light propagation in the Einstein-Born-Infeld black hole background.
Obviously, for $ Q = 0 $, they are just the null geodesic equations
for the Kerr black hole. The constant $K=0$ is the necessary and
sufficient condition for particles motion initially in the
equatorial plane to remain their. Any particle which crosses the
equatorial plane has $K>0$.
\paragraph{Effective potential.}
\begin{figure*}
\begin{tabular}{|c| c|}
\hline
\includegraphics[scale=0.8]{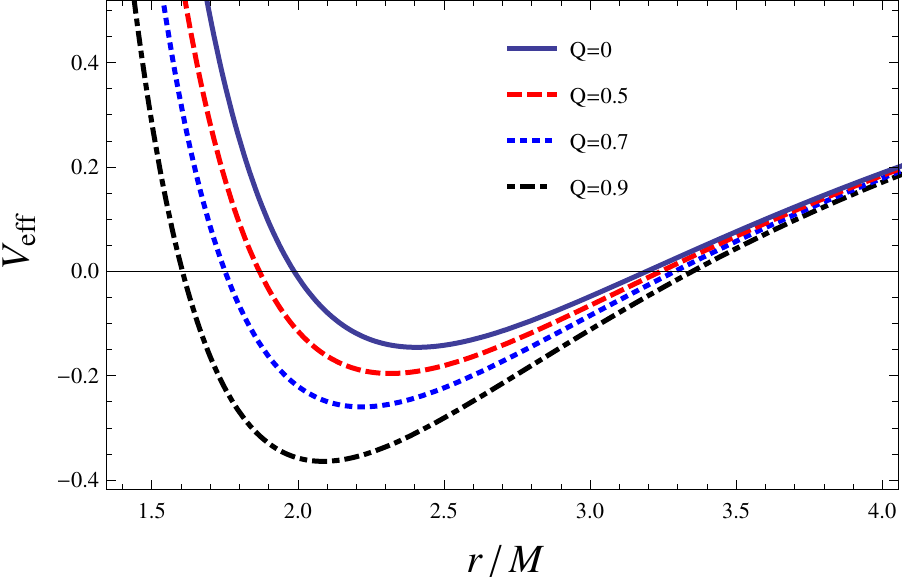}
&
\includegraphics[scale=0.8]{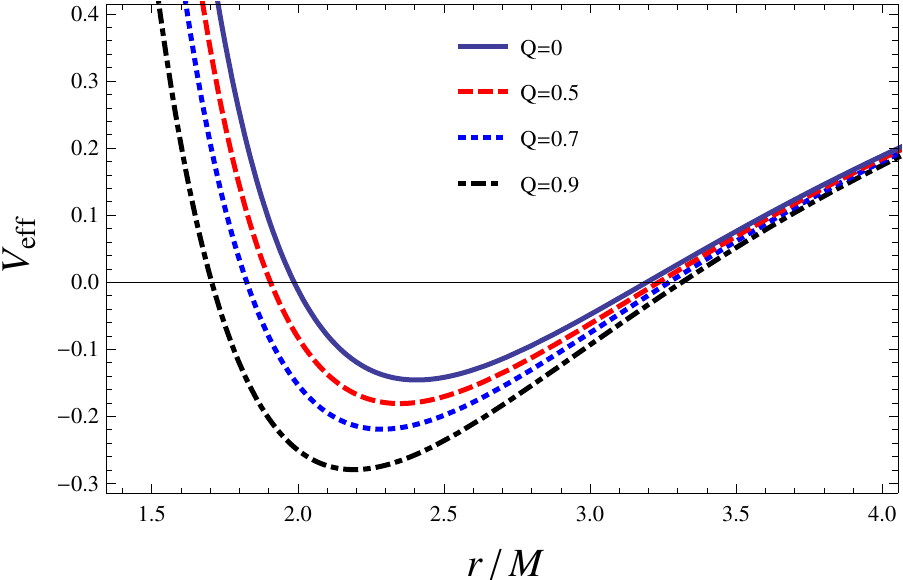} \\
\hline
\end{tabular}
 \caption{The radial dependence of effective potential $V_{eff}$ for the photon for the different values of electric charge $Q$.
  Left panel is for $\beta=0.05$ and $a=0.5$; right one is for $\beta=0.5$ and $a=0.5$.}\label{veff1}
\end{figure*}
%
%
\begin{figure*}
\begin{tabular}{|c| c|}
\hline
\includegraphics[scale=0.8]{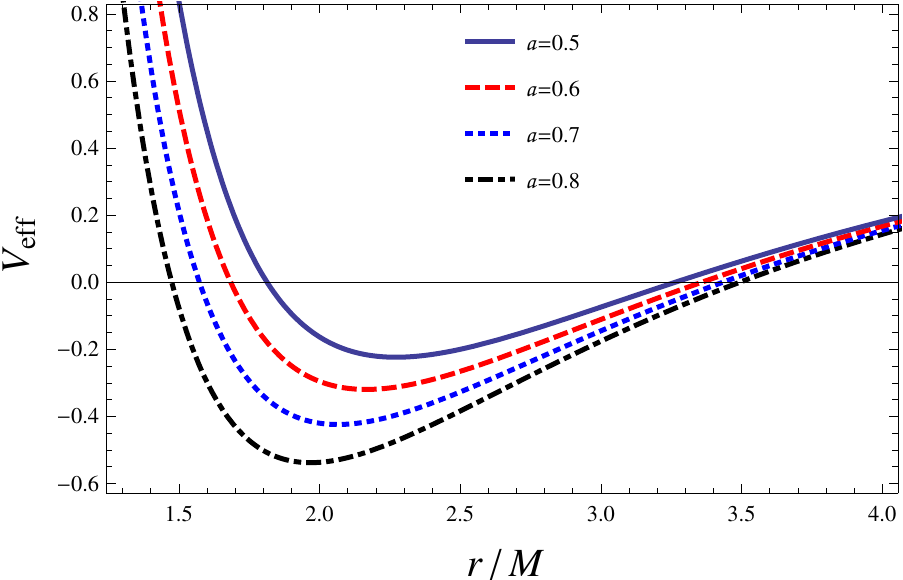}
&
\includegraphics[scale=0.8]{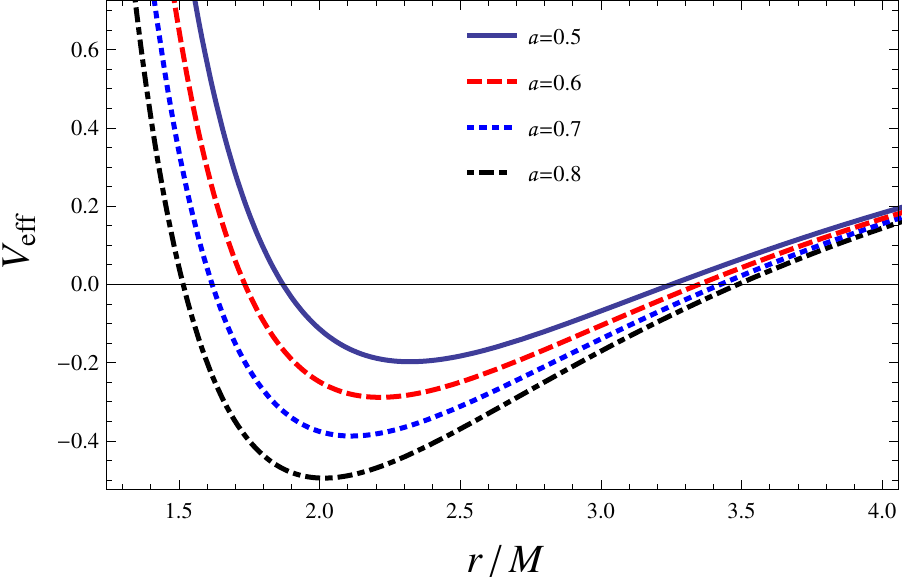} \\
\hline
\end{tabular}
 \caption{The radial dependence of effective potential $V_{eff}$ for the photon for the different values of rotation parameter $a$. Left panel is for $\beta=0.05$ and $Q=0.6$; right one is for $\beta=0.5$ and $Q=0.6$.}\label{veff2}
\end{figure*}
The discussion of effective potential is a  useful tool for
describing the motion of test particles. Further, we have to study
radial motion of photon for determining the black hole shadow
boundary. The radial equation for timelike particles moving along
geodesic in the equatorial plane ($\theta=\pi/2$) is described by
\begin{equation}
\frac{1}{2} \dot{r}^2 + V_{eff} = 0,
\end{equation}
with the effective potential
\begin{equation}
 V_{eff} =  -\frac{[E (r^2 + a^2) -La]^2 -\Delta  (L-a E)^2}{2 r^4}\ .
 \label{ve}
\end{equation}
From the last expression ~(\ref{ve}) one can easily get the
plots presented in Figs. \ref{veff1} and \ref{veff2}. There we have considered photon motion
around Einstein-Born-Infeld black hole for the different values of
electric charge $Q$ and parameter $\beta$. It is shown that with increasing
electric charge $Q$ or rotating parameter $a$ particle is going to
come closer to the central object.
\begin{figure*}
    \begin{tabular}{|c|c|}
    \hline
        \includegraphics[scale=0.8]{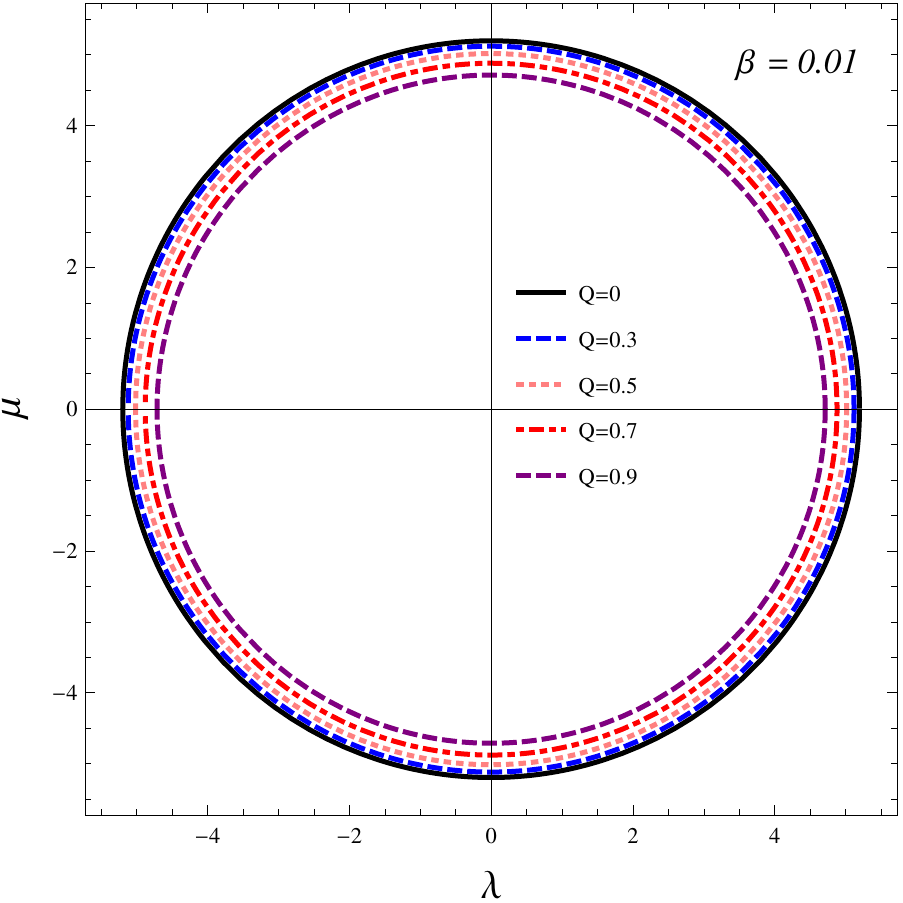}&
        \includegraphics[scale=0.8]{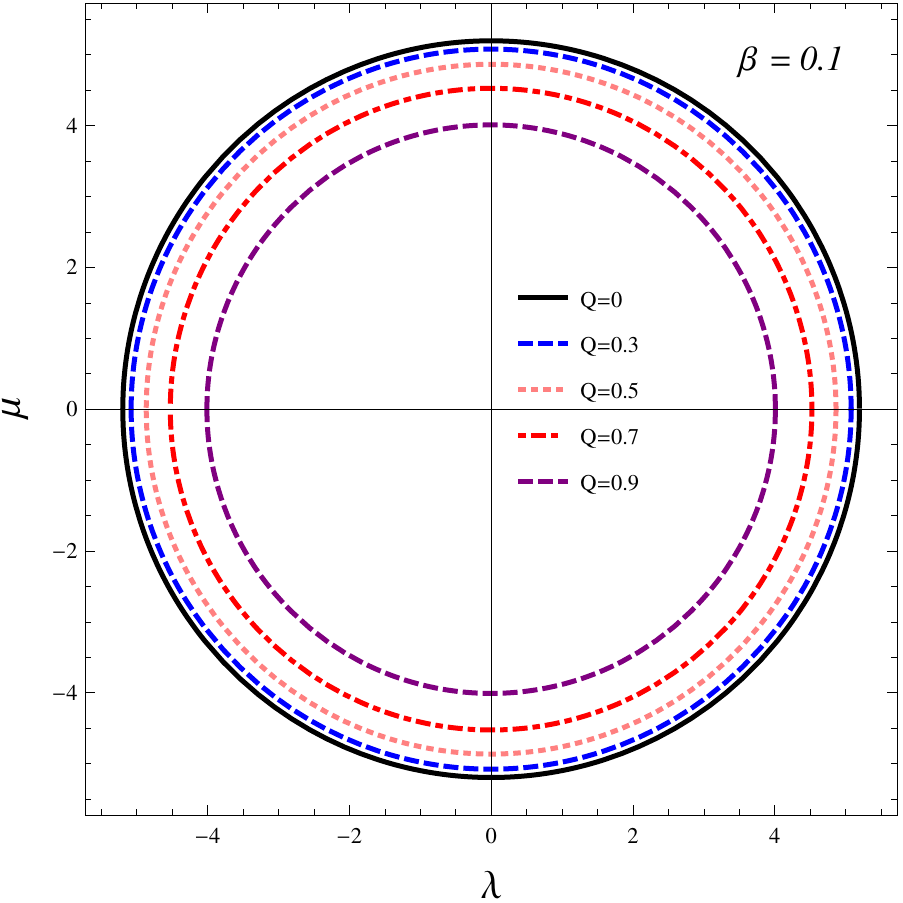}\\
  \hline \\
          \includegraphics[scale=0.8]{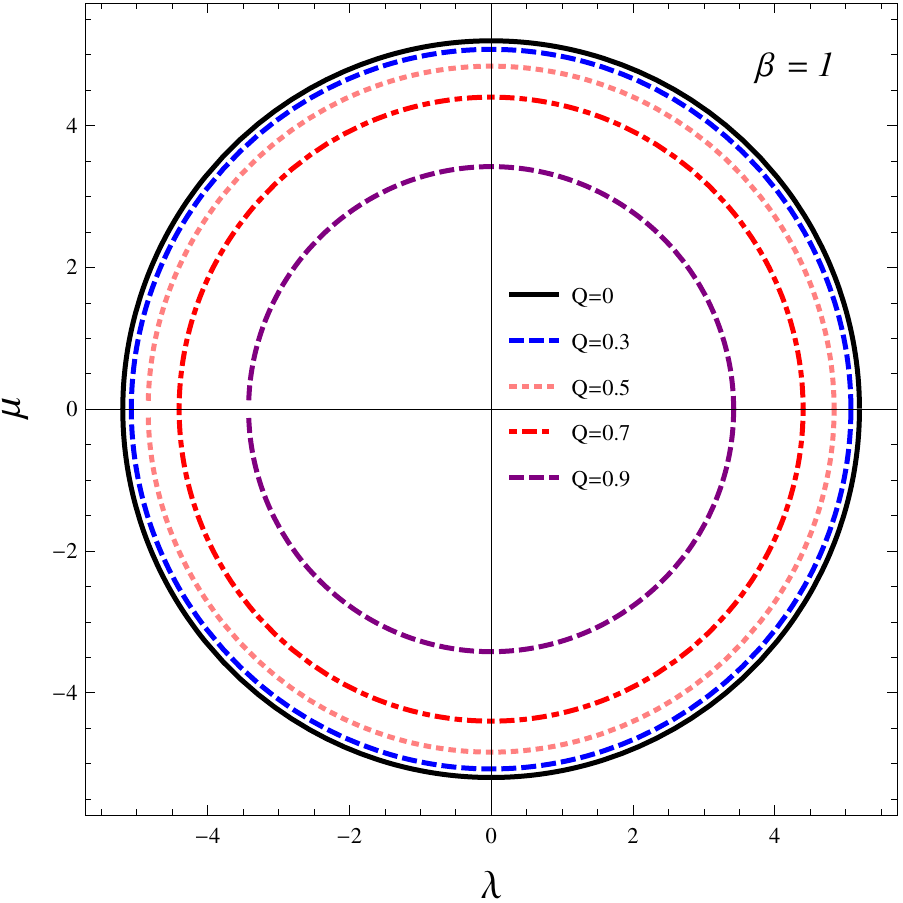}&
        \includegraphics[scale=0.8]{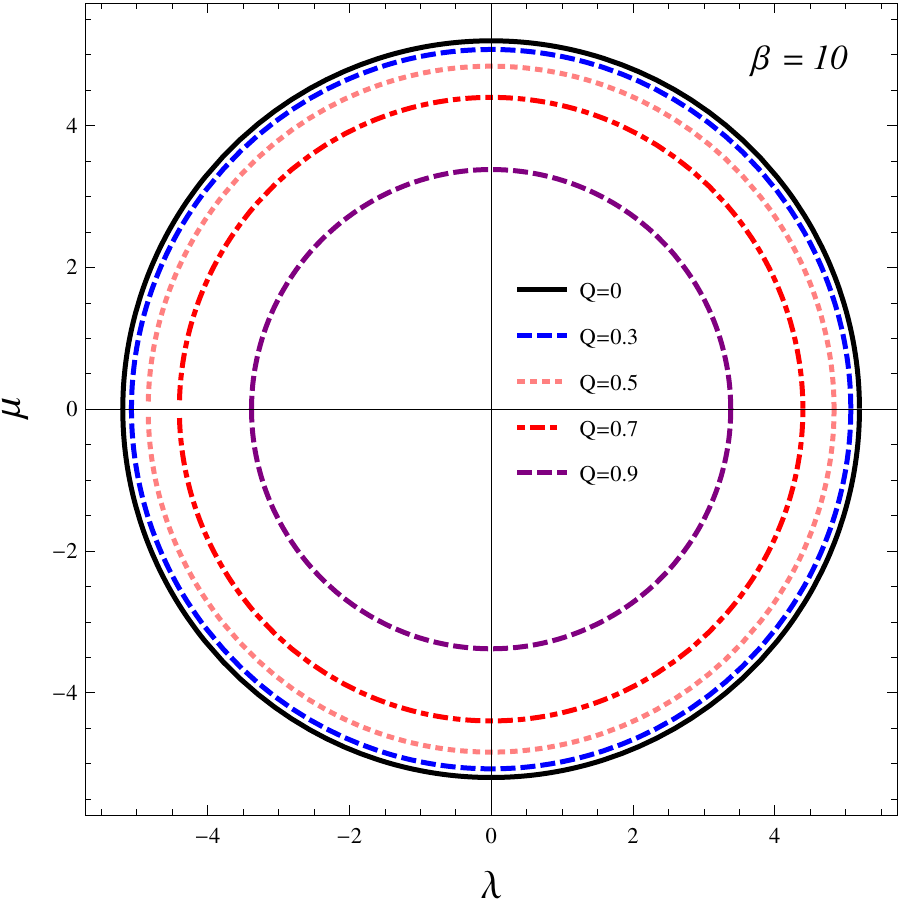} \\
                  \hline
    \end{tabular}
    \caption{Shadow of the black hole for
        the different values of electric charge $Q$. Top, left panel is for Born-Infeld parameter $\beta=0.01$. Top, right panel is for Born-Infeld parameter $\beta=0.1$. Bottom, left panel is for Born-Infeld parameter $\beta=1$. Bottom, right panel is for Born-Infeld parameter $\beta=10$ (with $M=1$ and $a=0$).}\label{shadi1}
\end{figure*}
\begin{figure*}
    \begin{tabular}{|c|c|}
    \hline
                \includegraphics[scale=0.8]{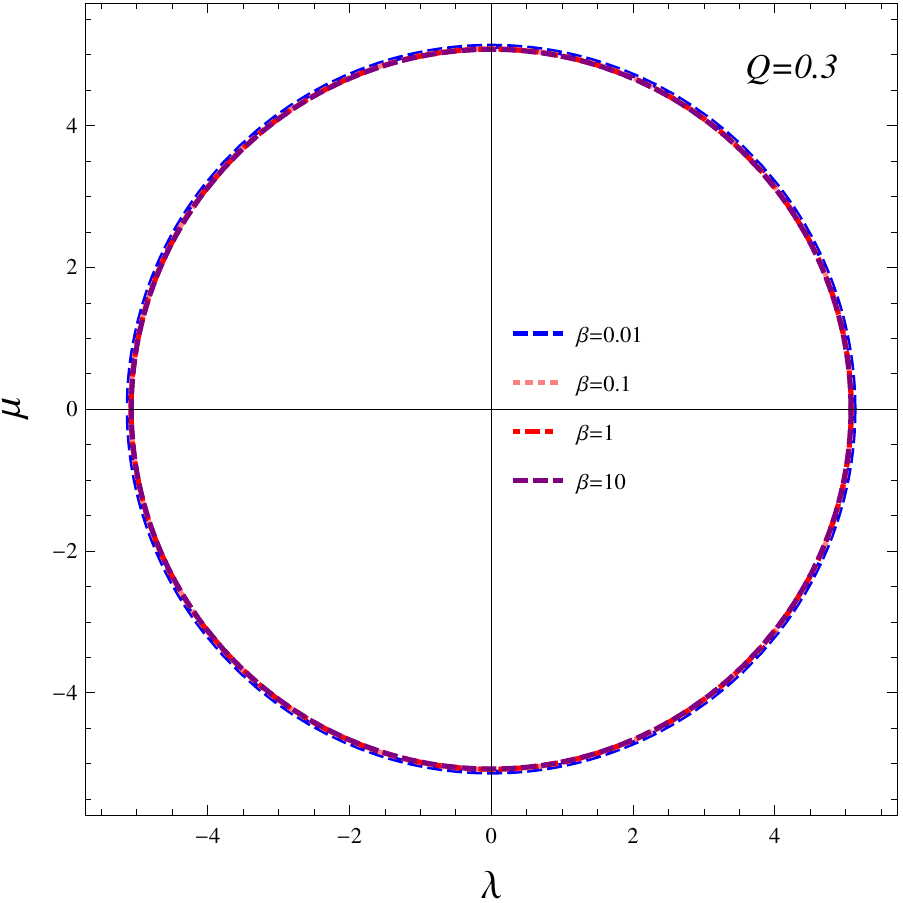} &
        \includegraphics[scale=0.8]{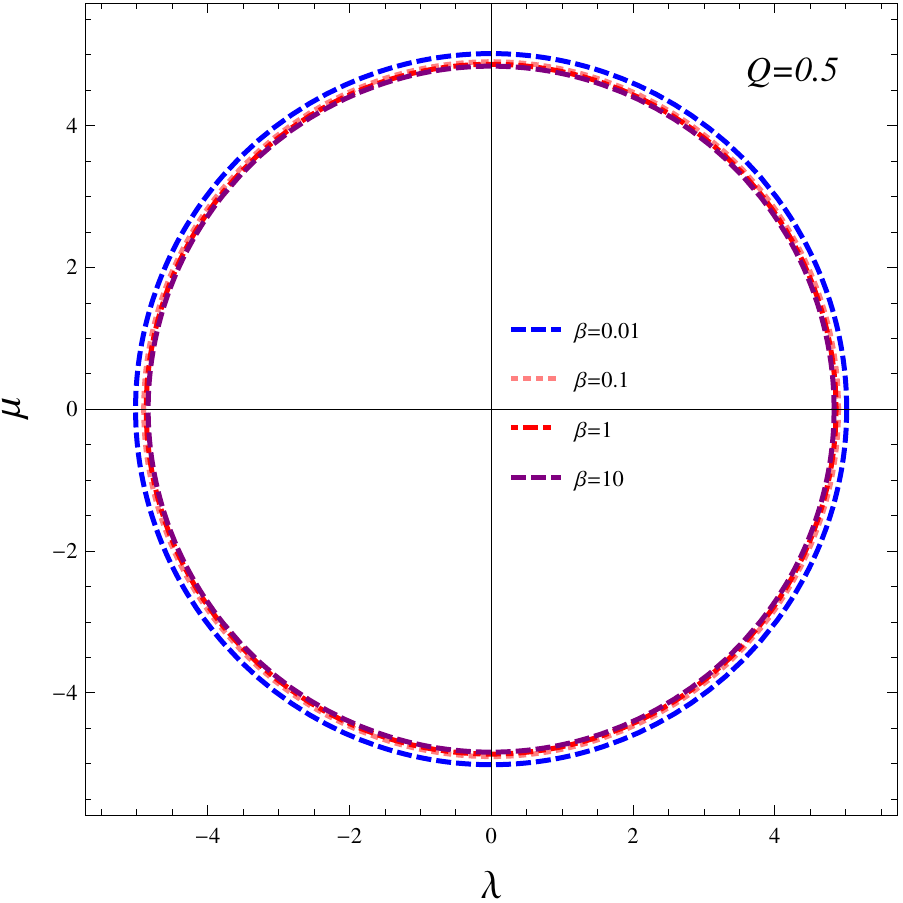}\\
\hline \\
        \includegraphics[scale=0.8]{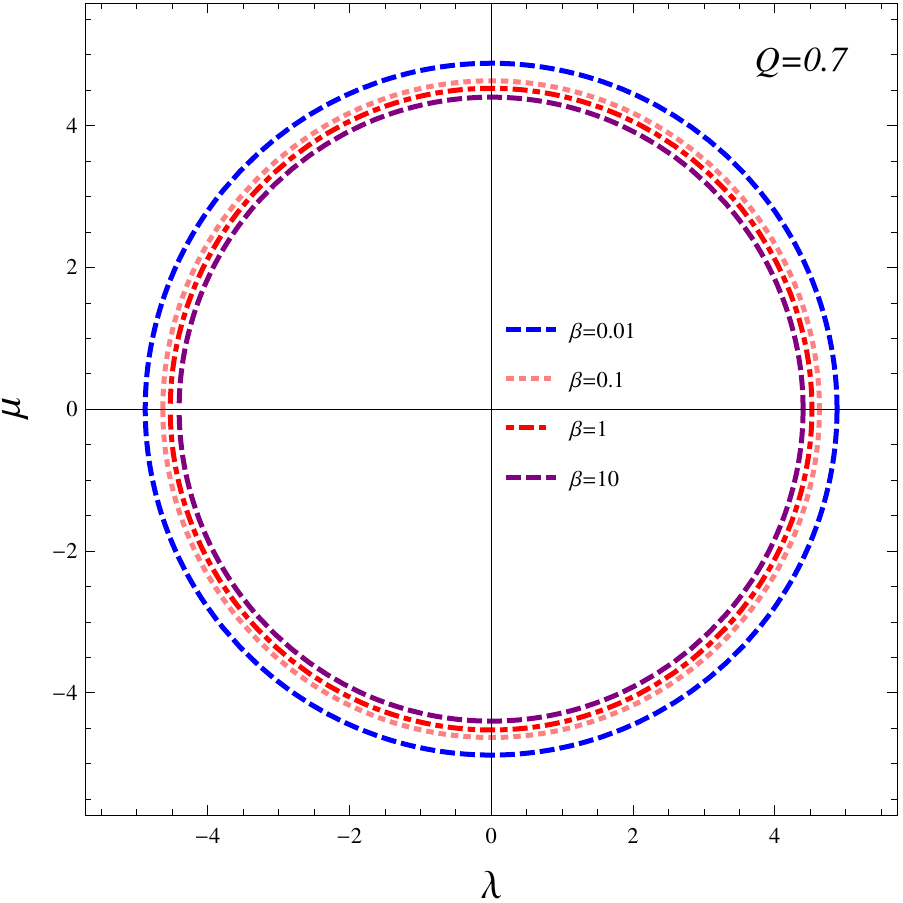}&
        \includegraphics[scale=0.8]{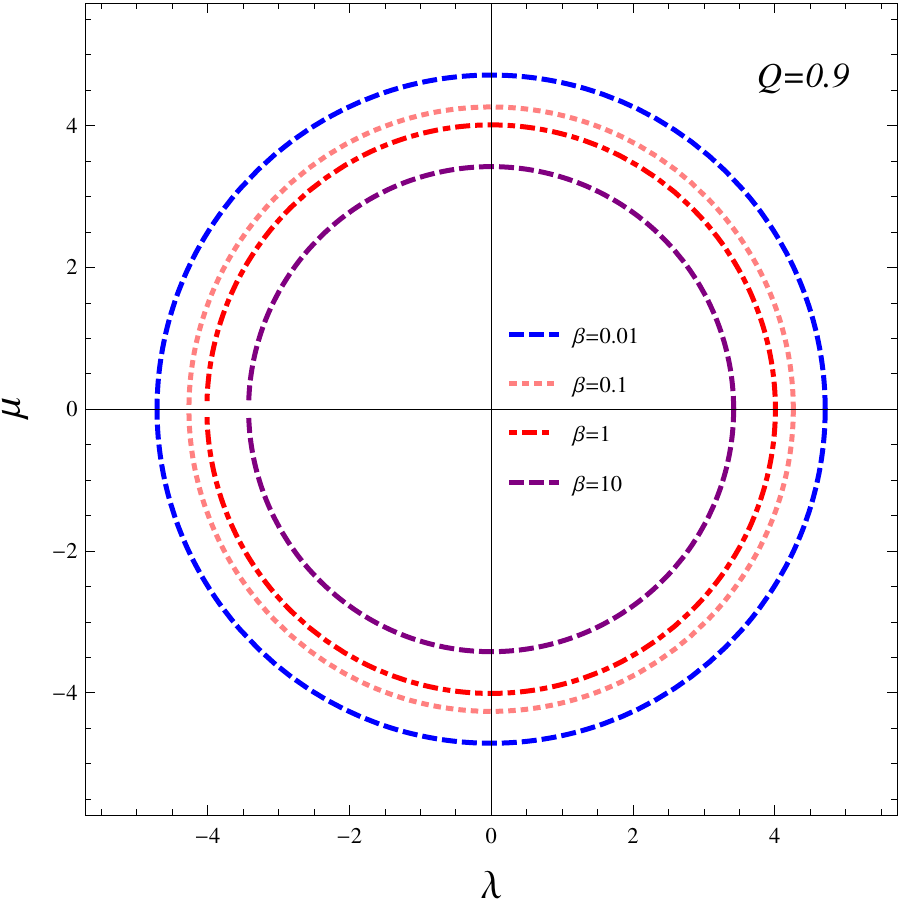}   \\
         \hline
    \end{tabular}
    \caption{Shadow of the black hole for
        the different values of Born-Infeld parameter $\beta$. Top, left panel is for electric charge $Q=0.3$. Top, right panel is for electric charge $Q=0.5$. Bottom, left panel is for electric charge $Q=0.7$. Bottom, right panel is for electric charge $Q=0.9$ (with $M=1$ and $a=0$).}\label{shadi2}
\end{figure*}
\begin{figure*}
\begin{tabular}{|c| c|}
\hline
\includegraphics[scale=0.8]{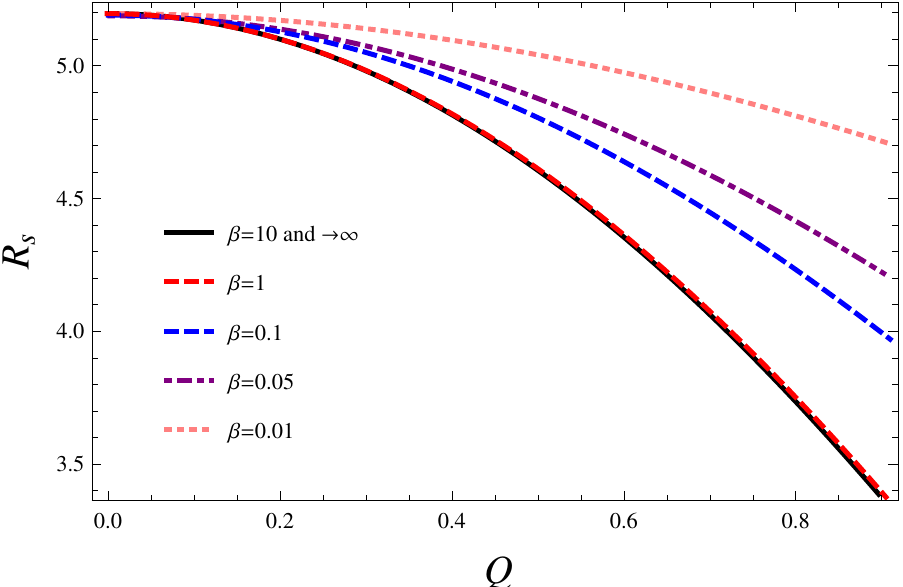}
\includegraphics[scale=0.8]{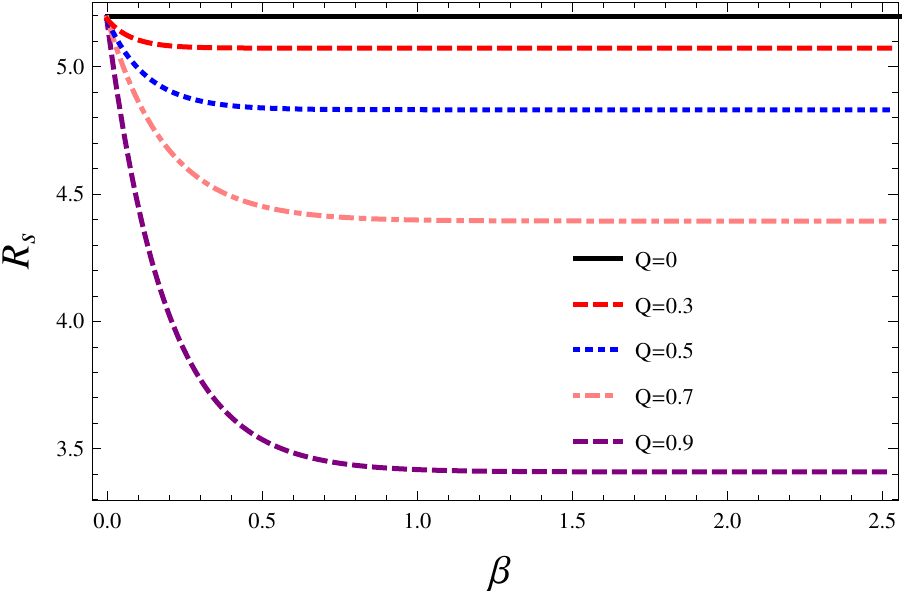} \\
\hline
\end{tabular}
 \caption{The dependence of observable radius of black hole shadow $R_{s}$ from the electric charge $Q$ and Born-Infeld parameter $\beta$. Left panel is showing graphs for the different values of  Born-Infeld parameter $\beta$. The right panel is showing graphs for the different values of electric charge $Q$.}\label{radiuss}
\end{figure*}
\section{Shadows of  Einstein-Born-Infeld black holes}
 Now, it is general belief that a black
hole, if it is in front of a bright background produced by far radiating object, will cast a shadow.
The apparent shape of a black hole silhouette is defined by the boundary of the
black and it was first studied by Bardeen \cite{JMB}. The ability of
very long baseline interferometry (VLBI) observation  has been improved
significantly at short wavelength which led to strong expectation  that
within few years it may be possible to observe the direct image of
the accretion flow around a black hole with a high resolution
corresponding to black hole event horizon \cite{vlbi}. This may allow
help us to test gravity in the strong field regime and investigate
the properties of black hole candidates. The VLBI experiments is
also looking for the shadow of a black hole, i.e. a dark area in
front of a luminous background \cite{JMB,shadow}.
 Hence, their is a significant attention towards study of
black hole shadow and it has become a quite active research field
\cite{Fara} (for a review, see \cite{bozzareview}). For the Schwarzschild  black hole the
shadow of the black hole is a perfect circle \cite{bozzareview}, and that enlarges in the case of Reissner-Nordstr$\ddot{o}$m black hole \cite{rn}
Here, we plan to discuss the shadow of the Einstein-Born-Infeld black hole, and we shall confine to non-rotating ($a=0$) case.
 It is possible to study equatorial orbits of photon
around Einstein-Born-Infeld black holes  via the  effective potential. It is generally known, that the photon
orbits are of three types: scattering, falling and
unstable ~\cite{Bambi}. \textit{The falling orbits} are due to the photons
arriving from infinity cross the horizon and fall down
into the black hole, they have more energy than barrier of the effective potential.
 The photons arriving from infinity move along
\textit {the scattering orbits} and come back to infinity, and
with
energy less than the barrier of the effective potential. Finally, the maximum value of the effective potential separates the
captured and the scattering orbits and defines unstable orbits of
constant radius (it is circle located at $r=3M$ for the
Schwarzschild black hole) which is responsible for the apparent
silhouette of a black hole. Distinct observer will be able to see only the photons scattered
away from the black hole, while those captured by the black hole
will form a dark region. If the black hole appears between a
light source and a distant observer, the photons with small impact
parameters fall into the black hole and form a dark zone in the
sky which is usually termed as black hole shadow.
We consider the following series expansion
\begin{eqnarray}
\frac{4Q^{2}}{3} F\left( \frac{1}{4},\frac{1}{2},\frac{5}{4},-\zeta^2(r) \right) \approx & &  \frac{4Q^2}{3} \left( 1- \frac{\zeta^2(r)}{10}\right)
\nonumber \\ & & + \mathcal{O}(\frac{1}{\beta^4})\ .
\end{eqnarray}
Also, all the higher order terms from
onwards have been dropped out from the series expansion of  $ F\left( \frac{1}{4},\frac{1}{2},\frac{5}{4},-\zeta^2(r)\right) $, which yields
\begin{eqnarray}
Q^2(r) & & \approx \frac{2\beta^{2}r^{4}}{3}\left( 1-\sqrt{1+\zeta^2(r)}\right) +
 \frac{4Q^2}{3} \left( 1- \frac{\zeta^2(r)}{10}\right)
\nonumber \\ & & + \mathcal{O}(\frac{1}{\beta^4})\ ,
\end{eqnarray}
and accordingly $\Delta$ is also modified and we denote the new $\Delta$ as $\Delta'$, which reads
\begin{eqnarray}
\Delta^{'} &=& -2+\frac{8Q^2 \zeta^2(r)}{15 r}+\frac{4Q^2}{3r \sqrt{1+\zeta^2(r)}}
\nonumber \\ & & +2r-\frac{8\beta^2 r^3}{3}\left(-1+\sqrt{1+\zeta^2(r)}\right)\ .
\end{eqnarray}\label{deltaprime}
Henceforth, all our calculations  are valid up to
$\mathcal{O}(\frac{1}{\beta^4})$ only.   An effective potential for
the photon attains a maximum, goes to negative infinity beneath the
horizon, asymptotically goes to zero at $r \rightarrow \infty$.  In
the standard Schwarzschild black hole, the maximum of the effective
potential occurs  at $r=3M$, which is also the location of the
unstable orbit and no minimum.  The  behavior of effective potential 
as a function of radial coordinate $r$ for different values of  parameter
$\beta$, rotation parameter $a$ and $Q$ are depicted  in Figs. \ref{veff1} and \ref{veff2}. It is observed that the
potential has a minimum which imply the presence of stable circular
orbits. The apparent shape of the black hole is obtained by
observing  the closed orbits around the black hole  governed by
three impact parameters, which are functions of  $E$, $L_{\phi}$ and
$L_{\psi}$ and the constant of separability ${\cal K}$.
The equations determining the unstable photon orbits,  in order to obtain the boundary of shadow of the black holes, are Eq.(\ref{bibh5}) or
\[R(r)=0=\partial R(r)/\partial r\]
which are fulfilled by the values of the impact parameters
\begin{equation}
\xi =  \frac{(r^2+a^2) \Delta^{'}-4 r \Delta}{\Delta^{'} a}\ ,
\end{equation}\label{ksiq}
and
\begin{equation}
\eta =  \frac{16\Delta r^2 a^2-\left((r^2+a^2) \Delta^{'}-4 r \Delta-a \Delta^{'}\right)}{(\Delta^{'} a)^2},
\end{equation}\label{etaq}
that determine the contour of the shadow.
whereas the parameter $\xi$ and $\eta$ satisfy
\begin{eqnarray}
&&\xi^2+\eta =\nonumber \\
&&\frac{4 A \Big(\frac{8Q^2 r_{0} \zeta^2(r_{0})}{15}-\frac{8Q^2 r_{0}}{3}+\frac{2Q^2 r_{0}}{3 \sqrt{1+\zeta^2(r_{0})}}+3r_{0}^2-r_{0}^3\Big)}{-1+\frac{4 Q^2 \zeta^2(r_{0})}{15r_{0}}+\frac{2Q^2}{3r_{0} \sqrt{1+\zeta^2(r_{0})}}+r_{0}-\frac{4}{3} \beta^2 r_{0}^3\left(-1+\sqrt{1+\zeta^2(r_{0})}\right)}
\nonumber \\ && +  \frac{60 \beta^2 r_{0}^4 B}{C} +O(a^2)\  \nonumber \\ \label{etaksi}
\end{eqnarray}
with
\begin{eqnarray}
A&=&-1+\frac{4Q^2 \zeta^2(r_{0})}{15 r_{0}}+\frac{2Q^2}{3r_{0} \sqrt{1+\zeta^2(r_{0})}} \nonumber \\ &&-r_{0}-\frac{4}{3} \beta^2 r_{0}^3\left(-1+\sqrt{1+\zeta^2(r_{0})}\right)
\end{eqnarray}
\begin{eqnarray}
B&=&6  Q^6  -14 \beta^2 Q^4 r_{0}^4+15 \beta^2 Q^2 r_{0}^5 - 10 \beta^4 Q^2 r_{0}^{8}+15 \beta^4 r_{0}^{9} \nonumber \\ &&+10 \beta^6 r_{0}^{12}-10 \beta^6 r_{0}^{12} \sqrt{1+\zeta^2(r_{0})}
\end{eqnarray}
\begin{eqnarray}
C&=&4 \beta Q^4\sqrt{1+\zeta^2(r_{0})}+ 20 \beta^5 r_{0}^8\left(-1+\sqrt{1+\zeta^2(r_{0})}\right) \nonumber \\ &&-5\beta^3 r_{0}^4 \Big(2Q^2-3r_{0}(r_{0}-1) \sqrt{1+\zeta^2(r_{0})}\Big)^2
\end{eqnarray}
The shadow of Einstein-Born-Infeld black holes may be determined through virtue of the above equation.
In order to study the shadow of Einstein-Born-Infeld black hole, it is necessary to introduce the celestial coordinates according to~\cite{Vazq}
\begin{equation}  \label{lambdaqb1}
\lambda=\lim_{r_{0}\rightarrow \infty}\left(
-r_{0}^{2}\sin\theta_{0}\frac{d\phi}{dr}\right)\ ,
\end{equation}
and
\begin{equation}  \label{muqb1}
\mu=\lim_{r_{0}\rightarrow \infty}r_{0}^{2}\frac{d\theta}{dr}\ ,
\end{equation}
The celestial coordinates can be rewritten as
\begin{equation}  \label{lambdaqb2}
\lambda=-\xi\csc\theta_{0} ,
\end{equation}
\begin{equation}  \label{muqb2}
\mu=\pm \sqrt{\eta + a^2 \cos ^{2}\theta_{0}-\xi \cot
^{2}\theta_{0}},
\end{equation}
and formally coincide with that for the Kerr black hole.
However in reality $\xi$ and $\eta$ are different for the Einstein-Born-Infeld
black hole. The celestial coordinate in the equatorial plane
($\theta_0=\pi/2$), where observer is placed, becomes
\begin{equation}  \label{lambdaqb2}
\lambda=-\xi,
\end{equation}
and
\begin{equation}  \label{muqb2}
\mu=\pm \sqrt{\eta}\ ,
\end{equation}
The apparent shape of the Einstein-Born-Infeld black hole shadow can be obtained by plotting $\lambda$ vs $\mu$ as
\begin{eqnarray}
\lambda^2+\mu^2=\xi^2+\eta \ ,
\end{eqnarray} \label{last}
which suggests that the shadow of Einstein-Born-Infeld black holes in ($\lambda,\mu$) space is a circle with radius of the quantity defined by the right hand side of equation (\ref{etaksi}).  Thus, the shadow of the black hole depends on the electric charge $Q$ and Born-Infeld parameter $\beta$ both.  These are depicted in Figs. \ref{shadi1} and \ref{shadi2} for the different values of these parameters.

In the limit, $\beta \rightarrow \infty$, the above expression reduces to
\begin{eqnarray}
\xi^2+\eta &=&  \lambda^2+\mu^2= \frac{2 r_{0}^2 (r_{0}^{2}-3)+4 r_{0} Q^2}{(r_{0}-1)^2}.
\end{eqnarray}\label{RN}
which is same as that for Reissner-Nordstrom black hole.
In addition, if we switch off electric charge $Q=0$, one gets expression for Schwarzschild case, which reads as
\begin{eqnarray}
\xi^2+\eta &=&  \lambda^2+\mu^2=\frac{2 r_{0}^2 (r_{0}^{2}-3)}{(r_{0}-1)^2}\ .
\end{eqnarray}\label{schwar}
In order  to extract  more detailed information from the shadow of the Einstein-Born-Infeld black holes, we must create the observables.  In general, there are two observable parameters as radius of shadow $R_{s}$ and distortion parameter $\delta_{s}$ ~\cite{hm}. For non-rotating black hole there is only single parameter $R_{s}$ which corresponds to radius of reference circle. From Figs. \ref{shadi1} and \ref{shadi2} one can get numerical value for the radius of black hole shadow which is clearly shown in Fig.\ref{radiuss}.

%
%

\begin{figure*}
\begin{tabular}{|c| c|}
\hline
\includegraphics[scale=0.8]{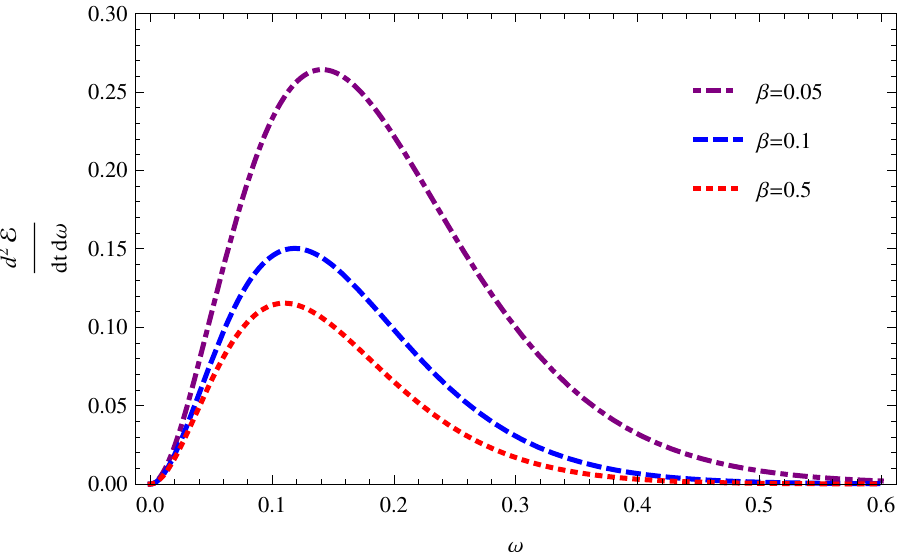}
&\includegraphics[scale=0.8]{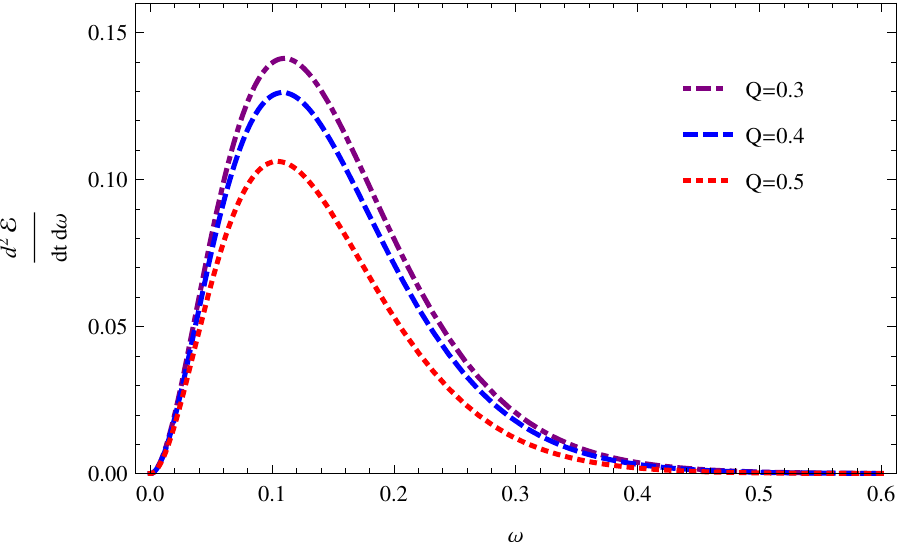} \\
\hline
\end{tabular}
 \caption{Energy emission of black hole in Einstein-Born-Infeld gravity. Left panel is for electric charge $Q=0.5$ and right panel is for Born-Infeld parameter $\beta=0.05$.}\label{enter}
\end{figure*}
%
At high energies the absorption cross section of a black hole has variation around a limiting constant value
and  for the distant observer placed at infinity the black hole shadow is responsible to its high energy absorption cross section.
For a black hole having a photon sphere, the limiting constant value coincides with
the geometrical cross section of the photon sphere \cite{Mtw}.
\section{Emission energy of  rotating Einstein-Born-Infeld black holes}
For completeness of our research here we investigate rate of the energy emission from the Einstein-Born-Infeld black hole with the help of
\begin{equation}
\frac{d^2E(\omega)}{d\omega dt}= \frac{2 \pi^2 \sigma_{lim}}{\exp{\omega/T}-1}\omega^3\ ,
\end{equation}
where $T=\kappa/2\pi$ is the Hawking temperature, and $\kappa$ is the surface gravity. At outer horizon the temperature $T$ is equal to
\begin{eqnarray}
T&=& \frac{2Q^4\sqrt{1+\zeta^2(r_{+})} -2 \beta^2 Q^2 r_{+}^4\left(1+2 \sqrt{1+\zeta^2(r_{+})}\right)}{12\beta^2 \pi r_{+}^5(a^2+r_{+}^2)\sqrt{1+\zeta^2(r_{+})} }\nonumber \\ &&+ \frac{3r_{+}^4 \beta^2 D}{12\beta^2 \pi r_{+}^5(a^2+r_{+}^2)\sqrt{1+\zeta^2(r_{+})} }\ , 
\end{eqnarray}\label{temp}
where
\begin{eqnarray}
D&=& -a^2 \sqrt{1+\zeta^2(r_{+})}+
\nonumber \\ &&r_{+}^2(\sqrt{1+\zeta^2(r_{+})}+2\beta^2 r_{+}^2(\sqrt{1+\zeta^2(r_{+})}-1))
\end{eqnarray}
The limiting constant $\sigma_{lim}$ defines the value of the
absorption cross section vibration of for a spherically
symmetric black hole:
\begin{equation}
\sigma_{lim} \approx \pi R_s^2\ . \nonumber
\end{equation}
Consequently according to \cite{Wei1} we have 
\begin{eqnarray}
\frac{d^2E(\omega)}{d\omega dt}=\frac{2\pi^3 R_s^2}{e^{\omega/T}-1}\omega^3\ . \nonumber
\end{eqnarray}
{The dependence of energy emission rate from frequency for the different values of electric charge} $Q$ {and parameter} $\beta$ is shown in Fig.~\ref{enter}.
{One can see that with the increasing electric charge} $Q$ {or parameter} $\beta$ {the maximum value of energy emission rate decreases, caused by horizon area decrease.}
\section{Conclusion}
In recent years the Born-Infeld action has received a significant attention to
the development of superstring theory, where it has been demonstrated  that the Born-Infeld action naturally  arises in
string-generated corrections when one considers an open superstring. This leads to interest in
extending the Reissner-Nordstrom black hole solutions in Einstein-Maxwell theory to the charged black hole solution in Einstein-Born-Infeld theory \cite{past}.
In view of this, we  have investigated the horizon structure of the charged rotating black hole solution in Einstein-Born-Infeld theory, and explicitly discuss the effect of the Born-Infeld parameter $\beta$ into event horizon and optical properties of black hole.  Further, this  rotating Einstein-Born-Infeld black hole solution
generalizes both Reissner-Nordstrom ($\beta \rightarrow \infty$ and $a=0$) and Kerr-Newman solutions ($\beta \rightarrow \infty$).  Interestingly, it turns out that for  given values of  parameters \{$M,Q, \beta$\} , there exists  $a=a_E$ for which the solution (\ref{bi})  can be an extremal black hole, which decreases with increase in the parameter $\beta$.   Further, we have also analyzed  infinite red-shift surfaces,  ergo-regions,  energy emission and Hawking temperature
of the rotating Einstein-Born-Infeld black hole. The Einstein-Born-Infeld black hole's horizon structure has been studied for the different values of  electric charge $Q$ and Born-Infeld parameter $\beta$, which explicitly demonstrates that outer (inner) horizon radius decreases (increases)  with increase in  the electric charge $Q$ and Born-Infeld parameter $\beta$.   We have done our calculations numerically as it is difficult to solve the analytical solution and found that the obtained results are different from the Kerr-Newman case due to non-zero Born-Infeld parameter $\beta$.
It is well known that a black hole can cast a
shadow as an optical appearance due to its strong gravitational
field. Using the gravitational lensing effect, we have also investigated the shadow cast by the non-rotating
($a=0$) Einstein-Born-Infeld black hole and  demonstrated that the
null geodesic equations can be integrated that allows us
investigate the shadow cast by a 
black hole which is found to be a
dark zone covered 
by a circle.  The shadow is slightly smaller and
less deformed than  that for its Reissner-Nordstrom counterpart. Further,
the shadow of the Einstein-Born-Infeld black hole is concentric
circle. In addition, the Born-Infeld parameter $\beta$  also changes the shape of the black hole's shadow.
In considered case of the nonrotating black hole only a radius of the shadow is
 observable parameter and from the numerical calculations we have easily got the value of the radius of the black hole shadow.
The effective potential for geodesic motion of the photon around rotating Einstein-Born-Infeld black hole
has been studied for the different values of the electric charge and spin parameter of the black hole.
With increasing either rotation parameter  or electric charge of black hole particle is moving closer to the central object.
Hence, circular orbit of the photon becomes closer to the center of rotating Einstein-Born-Infeld black hole.
It will be of interest to discuss energy extraction from rotating Einstein-Born-Infeld black hole, as the ergo-region is influenced by the Born-Infeld parameter and hence may enhance the efficiency of Penrose process. This and related work are the subject of forthcoming papers. Finally, in particular our results in the limit  $\beta \rightarrow \infty $, reduced exactly  to  Kerr-Newman black hole, and to Kerr black when $Q(r)=0$.
%
%
\section*{Acknowledgments}
BA thanks the TIFR and IUCAA for the warm hospitality during
his stay in Mumbai and Pune, India. This research is
partially supported by the projects F2-FA-F113, FE2-FA-F134 of the
Uzbekistan Academy of Sciences and by the ICTP through the OEA-PRJ-29 and OEA-NET-76 grants.
BA acknowledges the TWAS Associateship grant.
Support from the Volkswagen
Stiftung (Grant 86 866) is also acknowledged.
SGG thanks IUCAA for hospitality while part of the work was being done, and to  SERB-DST, Government of India for
 Research Project Grant NO SB/S2/HEP-008/2014.
%

\end{document}